\documentclass[aps,preprint,nofootinbib,superscriptaddress,showkeys,showpacs]{revtex4-1}
\usepackage{amssymb}
\usepackage{amsfonts}
\usepackage{amsmath}
\usepackage{amsthm}
\usepackage{epsfig}
\usepackage{color} 
\usepackage{subfigure}
\usepackage{lineno}
\usepackage{soul}
\usepackage{comment}
\usepackage{upgreek}

\newcommand{\fr}{\frac}

\begin{document}

\title{Online training of Echo State Networks to predict chaotic dynamical systems}

\author{A. Provata}
\affiliation{Institute of Nanoscience and Nanotechnology, National Center for Scientific Research ``Demokritos'', 15341, Athens, Greece}
\author{M. Syroukis}
\affiliation{Institute of Nanoscience and Nanotechnology, National Center for Scientific Research ``Demokritos'', 15341, Athens, Greece}
\author{J. Hizanidis}
\affiliation{Institute of Nanoscience and Nanotechnology, National Center for Scientific Research ``Demokritos'', 15341, Athens, Greece}
\affiliation{Institute of Electronic Structure and Laser, Foundation for Research and Technology-Hellas, Herakleio, 70013, Crete, Greece}
\date{\today}

\begin{abstract}
Echo State Networks (ESNs) are fundamentally important from the conceptual perspective of AI, because they provide simple and tractable frameworks for understanding temporal information processing, memory, and learning in dynamical 
systems. They are composed by an interconnected reservoir whose connectivity matrix remains constant over time and only the
output weights are updated during the training period.
In particular, ESNs employing online training allow the incremental adaptation of the output weights
as data is received, making them particularly suitable for applications in which new input data becomes available over time
(non-stationary environments). 
A key objective in the study of such learning structures is to determine which statistical properties of the reservoir enhance computational efficiency and maximize prediction performance.
To this end, we investigate the effects of reservoir topology on learning performance, considering 
 weight matrices generated from Erd\"os-R\'enyi random networks and Watts \& Strogatz small-world networks.
As test data we use the Lorenz and R\"ossler models in the chaotic regime and the FitzHugh-Nagumo model in the
limit cycle regime. 
In the case of random Erd\"os-R\'enyi reservoirs, our simulations show that best performances are obtained for
sparse reservoirs and especially when the link-density is near and immediately above
the critical point of mean-field bond percolation.
Watts \& Strogatz reservoirs are shown to demonstrate best performance also for small values of link-density,
compatible with the mean-field bond percolation when taking into account the specific nonlocal ring structure 
of the Watts \& Strogatz reservoirs. 
These studies on the influence of the reservoir architecture and hyperparameters on enhancing the accuracy of ESN predictions 
provide an efficient framework for real-time forecasting in dynamically evolving systems.

\end{abstract}


\maketitle
\section{Introduction}
Echo State Networks (ESNs)
 provide nowadays an efficient and powerful framework for processing and predicting temporal data 
and for modeling complex dynamical systems. They are simpler than conventional recurrent neural networks in the sense
that ESNs are
composed by only three layers, input layer, reservoir and output layer, of which only the output layer is trained;
the input weights and the reservoir connectivity matrix are kept fixed in time.  This
limited training results in  significantly lower training costs and in avoiding many of the optimization difficulties 
associated with gradient-based learning. Despite this simplicity, ESNs have demonstrated remarkable performance 
in tasks such as time series prediction, signal processing, and nonlinear control. 
They also provide an attractive framework for studying learning processes
using approaches adopted from the theory of complex systems, 
linking machine learning with nonlinear dynamics and network theory.

\par Echo State Networks \cite{jaeger:2001,jaeger:2002,jaeger:2007} together with their twin Liquid State Machines
(LSMs) \cite{maass:2002} were both introduced in the early 2000s to address learning as a result of
processing of input data by a nonlinear reservoir performing as an excitable medium. Since then, they have
gained considerable attention 
and are currently applied in learning tasks
 related to practical applications especially in the domain of complex
time series prediction \cite{racca:2021,trierweiler:2021,heyder:2022,gonzalez:2022,sharifi:2023,wang:2025,mandal:2025,yadav:2026}.
Apart from ESNs, reservoir computing has been recently employed within a broader 
category of machine learning approaches 
for time series predictions and chaotic attractor reconstruction, using either
 artificial neural networks or the dynamics of real-world physical systems 
\cite{pathak:2018,lu:2018,vlachas:2020,wikner:2024,li:2013,schumacher:2015,pandey:2020,arun:2024,arun:2025,thorne:2025,xu:2025,xu:2025_2,amann:2026,baxevanis:2026,rathor:2026}.

\par The performance of ESNs and LSMs is strongly influenced by the choice of the reservoir hyperparameters, 
and the input and output weight vectors. Parameters such as the reservoir size and connectivity,  the 
scaling of the input data and the leaking rate (governing the balance between previous and current reservoir states), 
directly affect
the learning process. Since the ESN reservoir internal weights remain fixed during training, selecting 
appropriate hyperparameters is essential for achieving high prediction accuracy, efficient representation of 
complex dynamics, and best predictive power. Understanding how these hyperparameters influence performance 
also provides valuable insights into the underlying mechanisms responsible for learning in reservoir computing
in particular and in AI, in general.

\par Investigating the role of hyperparameters is equally important from a practical perspective. 
Large and densely connected reservoirs generally offer richer dynamics but require higher computational 
and memory costs, whereas smaller or sparser reservoirs can substantially reduce these requirements 
while maintaining comparable performance. Identifying optimal trade-offs between hyperparameters can 
therefore lead to more efficient reservoir designs, particularly for real-time applications. A systematic analysis of hyperparameter effects not only facilitates the development of guidelines for reservoir design but 
also addresses scalability, robustness, and energy efficiency of reservoir computing systems 
across a broad range of applications.

\par A practical class of ESNs is the ``Echo State Networks with OnLine Training'' (OLT\_ESNs) 
as described in the 2002 seminal paper by H. Jaeger \cite{jaeger:2002}.
Typical, offline Echo State Networks form state matrices during the training and 
 construct the output weights only once, at the end of the training period. In contrast,  
OLT\_ESNs are designed to accept sequential data and to train  the output 
weights online, as new data points enter the training phase \cite{jaeger:2002}.
One of the main advantages of using online training in Echo State Networks is the simplicity 
in the updating of the output weights, which evolves as a step-by-step adaptation process in time.
 Another advantage of online reservoir training over offline training
 is that there is no need to invert the reservoir state matrix, which can be a difficult task 
often leading to singularities and to huge output weights.
On the other hand, a significant
 disadvantage of online training is that it is more time consuming, as the algorithm
requires updating the output weights every time a new training
data point becomes available for training, while in the case of offline training
only the reservoir state matrix is updated during training
 and the matrix inversion takes place only once, at the end of the training
period. Despite this disadvantage, the online ESN training remains a straightforward approach,
promotes sequential training allowing to adjust the weights easily as new training
data points become available, and avoids repeated inversions 
of the large, updated state matrix
whenever additional training data points are incorporated. In the
present study, the OLT\_ESN method will be applied 
in a step-by-step approach in order,
on the one hand, to allow for an in-depth understanding of the learning process and, 
on the other hand, to  access  the ESN prediction performance with respect to the hyperparameters related to
the reservoir connectivity.

\par More specifically, we concentrate on the influence of the reservoir connectivity on its performance, 
considering the reservoir from the perspective of complex network theory. 
Our aim is not to identify the hyperparameters which produce absolute best performance but to highlight 
the influence of the reservoir structural complexity on its prediction power. 

\par In particular, we compare reservoirs formed as Erd\"os-R\'enyi (ER) 
random networks
\cite{erdos:1959,barabasi:2016} to Watts \& Strogatz (WS) small-world networks \cite{watts:1998,barabasi:2016} 
with respect to their  efficiency in predicting the next elements in chaotic and nonlinear periodic time  series. 
As test cases we use the  Lorenz, 
R\"ossler and FitzHugh-Nagumo models.   Varying the network connectivity in ER networks
and the rewiring probability in WS networks and keeping all other hyperparameters fixed,
 we show that 
best performances are achieved when the network operates with connectivity around 
and beyond the bond percolation threshold.
We attribute this to the ability of the reservoir to diffuse information throughout its entire volume,
when it operates on or above the bond percolation
threshold. In the opposite case, for very sparse reservoirs
with connectivity below the percolation threshold, the information may remain localized and does not  spread through
the entire reservoir.

\par The present study is organized as follows. In the next section, 
the main background material needed
for this study is recapitulated. Namely, in Sec.~\ref{sec:ESN} the main ideas 
on Online Echo State Networks and the OLT\_ESN algorithm are briefly summarized. 
In Sec.~\ref{sec:reservoir-connectivity}, 
the different models of the reservoir connectivity are described. 
In Secs.~\ref{sec:Lorenz}, ~\ref{sec:Roessler} and ~\ref{sec:FHN}, 
three nonlinear dynamical systems are proposed for training and testing the ESN 
prediction algorithm: these
are the Lorenz, R\"ossler and FitzHugh-Nagumo models, respectively. 
In Sec.~\ref{sec:Results} the main
findings of this study are presented,
with the prediction results for Erd\"os-R\'enyi random networks
 discussed in Sec.~\ref{sec:predicitions-ER}
 and those for Watts \& Strogatz small-world networks in Sec.~\ref{sec:predicitions-WS}. 
In both cases, the parameter regions
where the reservoirs demonstrate best performance are reported and discussed 
in connection with the reservoir structural properties.
In the concluding section, Sec.~\ref{sec:conclusion}, the main findings of 
this study are recapitulated and open problems are discussed. In Appendix~\ref{appendixA},
the OLT\_ESN algorithm is presented in detail for clarifying all its steps.
Appendix~\ref{appendixB} is devoted to 
exemplary computations for determining appropriate sizes of the test sets. 
In Appendix~\ref{appendixC}, the effects of limited training data sets
on predictions of the Lorenz
and FHN dynamics are discussed.

\section{The Echo State Network and the Training Data Sets }
\label{sec:models}
In this section the structure of the Echo State Network is discussed. First,
the main features of the OLT\_ESN algorithm  are briefly
described in Sec.~\ref{sec:ESN},
while the full detailed structure of the algorithm is provided in Appendix~\ref{appendixA}.
 The different types of reservoir connectivity 
used in this study are presented in
detail in Sec.~\ref{sec:reservoir-connectivity}.
In Secs.~\ref{sec:Lorenz}, ~\ref{sec:Roessler} and ~\ref{sec:FHN}
the three dynamical systems are briefly introduced; the data series derived
from these models
are used as representative training and testing data sets
 to assess the predictive power
of the Echo State Network for the different types of reservoir connectivity. 
For each one of the three dynamical systems used to produce the 
training-test data sets, the working parameters are provided in the corresponding
sections.

\subsection{The Echo State Network with online training}
\label{sec:ESN}
\par All ESNs consist primarily of a reservoir composed by $N$ interconnected nodes. 
All nodes are linked
via a connectivity matrix ${\bf W}=\left[ w_{ij}\right]$ of dimensionality 
$N \times N$, with $w_{ij}$ representing the connectivity weight between
pre- and post-synaptic nodes, $j$ and $i$, respectively. The 
structural properties of matrix ${\bf W}$ are  
discussed in Sec.~\ref{sec:reservoir-connectivity} and, in most cases related to applications,
sparse connectivity matrices are used. The connectivity matrix is randomly initialized
 with positive and negative
weights representing excitatory and inhibitory interactions, respectively, and
remains unchanged throughout the entire training and testing stages. 

\par The training data is fed into the reservoir via an input weight vector, 
$\overrightarrow{W}^{\rm in}=\left( w^{\rm in}_1, w^{\rm in}_2,\dots , w^{\rm in}_N, \right)$, of size $N$. 
Namely, the element $w^{\rm in}_i$ provides the weight that scales the data fed to node $i$,
 $i=1,2, \cdots N$.  The input weights are randomized in the interval $(-1,1)$ using typically 
a uniform distribution  and also remain constant during the entire process.

\par In ESNs, learning is achieved only by adjustments on the output weights,
represented by the corresponding vector $\overrightarrow W^{\rm out}$.
The output weights $w^{\rm out}_i$, $i=1, \dots , N+1$,
are initially set to 0, but they change drastically and repeatedly during the
course of the online training. We note that the output weights 
contain one more component than the
number of nodes. This is called the bias node and is added in order to allow the readout layer to 
take into account non-zero offsets of the data, see further explanations  in Appendix~\ref{appendixA}.
When online training is assumed, as in the present study, 
every time a new data point is fed to the reservoir, 
new adjustment is applied to the output weights and these are used to predict the next
data element.

\par The state of the reservoir at every time step, $n=1, \dots , K_{\rm training}$, 
of the training period
is expressed as a vector $\overrightarrow X(n)$, 
where $x_i(n)$ is the value of the node $i$, $i=1, \dots , N$, of the reservoir at
time step $n$. The state values of all nodes of the reservoir are initiated
 to zero for all $i$ and  are sequentially updated during the entire process.

\par A brief description of the OLT\_ESN algorithm at the 
Training and the Testing stages follows below (for full details,
please refer to Appendix~\ref{appendixA}). 

\par At the Training Stage, at every time step $n$,  one
data point  is fed to all the nodes of the reservoir scaled with the appropriate
weights $w^{\rm in}_i$. Each node $i$ also receives incremental contributions  from all
other reservoir nodes $j$ with preselected weight $w_{ij}$. The sum is then passed 
through a nonlinear activation function. In the present study the hyperbolic tangent is
used as the activation function. Often, at this stage a new
augmented state, $z_i$, $i=1, \cdots N+1$, is prepared to add extra freedom to the reservoir.
Then, the next predicted value is computed using the output weights that have been obtained 
in the previous training step of the algorithm. Comparison of the predicted and the 
real data point normally 
results in an error. This error is used for updating the output weights.
In the present study, the ``Widrow-Hoff LMS algorithm''
is adopted for the output weight updates \cite{widrow:1960,koford:1966}. The Training Stage
ends when all $K_{\rm training}$ elements of the training data set have been fed to the
reservoir and the output weights have been adapted accordingly.

\par At the end of the Training Stage, the output weights are fixed and are not updated
during the Testing Stage. At every time step of the Testing Stage, 
a data value $I^{\rm test}(n)$ from test data
is fed to all nodes scaled with the appropriate weights $w^{\rm in}_i$.
As in the Training Stage, each node $i$ also receives incremental contributions from all
other reservoir nodes $j$ with the preselected weight $w_{ij}$ and the total
sum is then passed 
through the activation function. The next predicted value is computed 
using the fixed output weights that have been computed at the end of the Training Stage. 
Comparison of the predicted and the 
real data point may still result in an error. This process is repeated over the entire
testing set and all accrued
error values are summed and normalized. The resulting Normalized Root Mean Square
Deviation (NRMSE) value (see formula Eq.~\eqref{eq17} in Appendix~\ref{appendixA}) 
gives the overall 
predictive power of the OLT\_ESN algorithm.

\par The exact detailed algorithm with all formulae used in the present study is 
provided in Appendix~\ref{appendixA} together with the description of all parameters
and variables. 

\subsection{The reservoir connectivity}
\label{sec:reservoir-connectivity}
One of the main factors that significantly influence the efficiency of the ESN algorithms is the structure
of the reservoir and its link density. 
In most recent studies the reservoir is modeled as a random network with a sparse connectivity
and the density of random links is set to approximately 1\% based on empirical studies.  In the present work
we pay attention to the reservoir properties and we test different cases corresponding
to well studied universal models of complex networks \cite{barabasi:2016,newman:2010}. These are: 
\begin{enumerate}
\item {\bf Erd\"os-R\'enyi (ER) reservoirs} \cite{barabasi:2016,erdos:1959}: 
In this case, the reservoir is a random network composed of $N$ nodes
linked with a probability, $p \in \left[0, 1 \right] $. In this study we  explore the efficiency of the
reservoir as a function of both $N$ and $p$.
\item {\bf Watts \& Strogatz (WS) reservoirs} \cite{barabasi:2016,watts:1998}: In this case, the reservoir is a k-ring, that is a network where
 every node is linked to $k/2$ elements to its left and $k/2$ elements to its right. Moreover,
short-cuts are introduced on the k-ring in the sense that with probability $p_{\rm WS}$, each link is 
cut and pasted to a random position in the network. The efficiency of the WS
reservoir is here explored as a function of both $k$ and $p_{\rm WS}$.
\item {\bf Barabasi-Albert (BA) reservoir} \cite{barabasi:2016,albert:2002}: 
In this case the reservoir of size $N$ is composed dynamically using
the Barabasi-Albert generation method. Namely, starting with a small number of randomly connected nodes,
at every time step we insert in the system a new node containing $m$ links. The new node attaches preferentially
to nodes of high connectivity according to 
the rule of preferential attachment:  If $i$ is one of the already
 existing nodes in the BA network and has degree $s_i$, 
the new incoming node will attach to node $i$ 
with probability  $q_i={s_i}/{\sum_{j=1}^N{s_j}}$. In the BA network, hub nodes are 
developed and the degree distribution
follows a power law \cite{barabasi:2016,albert:2002}. 
\end{enumerate}

The idea of using the above connectivity models stems from medical imaging studies of brain connectivity. 
Although there is broad agreement that the brain connectivity is not random, 
many studies have reported small-world connectivity in the brain 
\cite{bassett:2006,muldoon:2016,hilgetag:2016,liao:2017,han:2025}, 
while others have identified scale-free connectivity \cite{eguiluz:2005,lee:2010,expert:2011,ribeiro:2021,jones:2023}.  In the present work, we investigate the predictive performance
of a reservoir in the case of random (Erd\"os-R\'enyi) connectivity,
which is the connectivity most commonly used in reservoir computing,
 and compared it to that of a small-world (Watts \& Strogatz) network. The case of scale-free (Barabasi-Albert) connectivity  is left for future studies.

\subsection{Data: The Lorenz model}
\label{sec:Lorenz}

\par To keep the OLT\_ESN algorithm as simple as possible in order to better understand
the prediction process, we use 1D sequential data
in the form of time series produced by nonlinear dynamical systems.
As a first example, we use the sequence issued from the numerical integration of the chaotic  Lorenz 
model producing the well-known butterfly effect. 
The model has three variables, $x$, $y$ and $z$ and
its dynamics is given by the following ordinary differential equations \cite{lorenz:1963}:
\begin{subequations}\label{eq21}
\begin{align}
 \frac{dx}{dt} &=\sigma \left( y-x \right), 
      \label{eq21a} \\
 \frac{dy}{dt} &=x \left(\rho-z \right) -y, 
      \label{eq21b} \\
 \frac{dz}{dt} & =x y - \beta z.
      \label{eq21c}
\end{align}
\end{subequations}

\par For training and testing the OLT\_ESN algorithm, we prepare a time series as
a data set of the $x$-component of the Lorenz model.
To ensure that the model operates in its full chaotic regime, we use the following
parameter set: $\sigma=10.0$, $\rho =28.0$ and $\beta =8/3$.
For these parameters the chaotic Lorenz attractor does not cover the full 3D space
but only a fractional part of it, with fractal dimensions $d_f=2.06$.
 See a typical phase space plot of the Lorenz attractor for the above parameter values in Fig.~\ref{fig01}(a).

\subsection{Data: The R\"ossler model}
\label{sec:Roessler}
As a second example of training and testing chaotic time series data, 
we use the R\"ossler dynamical system
\cite{roessler:1976}.
This model  also has three variables $x$, $y$ and $z$ and its dynamics is described by the following
 nonlinear differential equations:

\begin{subequations}\label{eq22}
\begin{align}
\label{eq22a}
\frac{d x}{dt} & = y - z \\
\label{eq22b}
\frac{d y}{dt} &= x + a y, \\
\label{eq22c}
\frac{d z}{dt} &= b + z (x - c).
\end{align}
\end{subequations}

\par As in the case of the Lorenz model, here also  we prepare the
$x$-component data set by numerically integrating
the R\"ossler model. The parameter values to produce
fully developed chaos are: $a = b = 0.2$ and $c=5.7$
and a typical phase space plot of the attractor
is displayed in Fig.~\ref{fig01}(b). For these parameter values, the R\"ossler attractor presents 
a fractal dimension $d_f=2.01$, slightly lower than the Lorenz attractor.

\subsection{Data: The FitzHugh-Nagumo model}
\label{sec:FHN}
\begin{figure*}[h]
\includegraphics[width=0.36\linewidth]{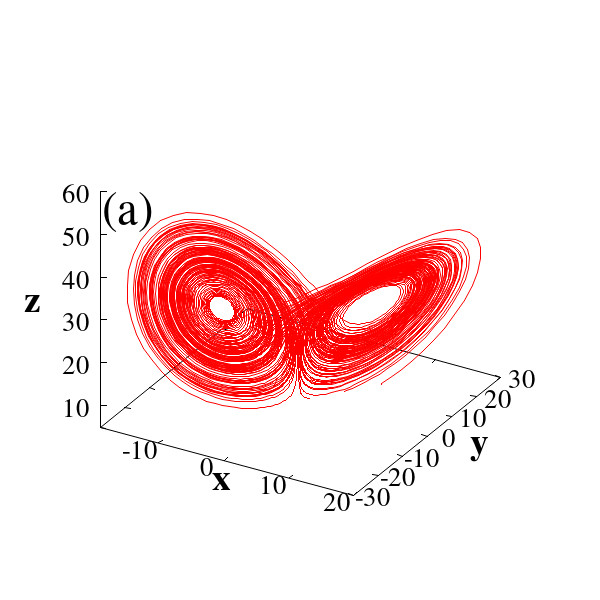}
\includegraphics[width=0.36\linewidth]{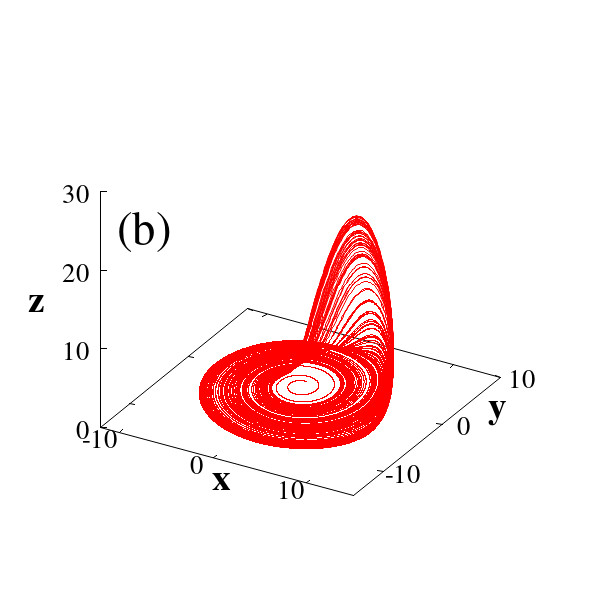}
\includegraphics[width=0.26\linewidth]{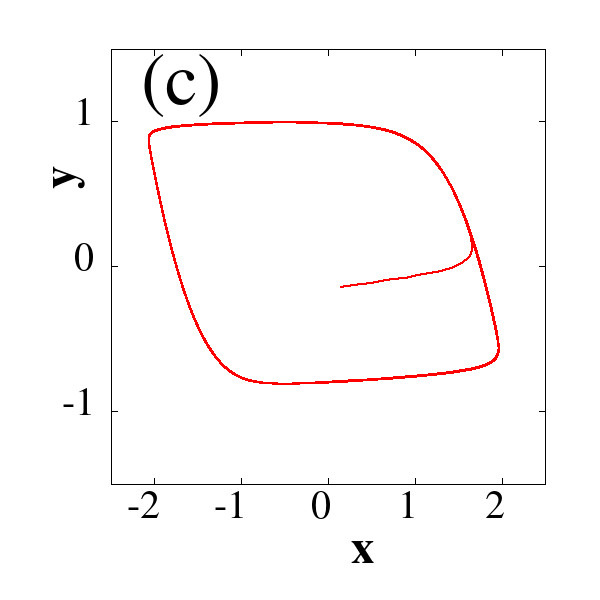}
\caption{(Color online) Phase space plots of:  a) the 3D chaotic Lorenz attractor, 
b) the 3D chaotic R\"ossler attractor 
and c) the 2D FitzHugh-Nagumo nonlinear oscillator. For the three models,
the parameter values used in the 
corresponding Runge-Kutta integrations are described in Secs.~\ref{sec:Lorenz}, ~\ref{sec:Roessler} 
and ~\ref{sec:FHN}, respectively.}
\label{fig01}
\end{figure*}

This is a model describing a nonlinear oscillator in 2 dimensions that was proposed by Richard FitzHugh
in 1961 \cite{fitzhugh:1961}
as a simplification of the Hodgkin-Huxley model describing the dynamics of a spiking neuron
\cite{hodgkin:1952}. Independently, 
the following year Jinichi Nagumo and co-workers created the equivalent circuit \cite{nagumo:1962}.
The equations describing the FHN systems are:
\begin{subequations}\label{eq23}
\begin{align}
\label{eq23a}
\varepsilon \frac{d x}{dt} & = x - \fr{x^3}{3} - y + Q \\
\label{eq23b}
\frac{d y}{dt} &= x + \upalpha,
\end{align}
\end{subequations}
where $x$ and $y$ variables denote the activator and inhibitor potentials,
respectively \cite{fitzhugh:1961,nagumo:1962, rontogiannis:2021}. 
The $x$ component of the FHN trajectory will be used for the generation of the
training and testing data sets.
 In Eqs.~\eqref{eq23}, $\varepsilon>0$ is a small parameter characterizing the
 timescale separation between the activator 
and inhibitor evolutions. 
 We fix it at $\varepsilon = 0.05$ throughout this study to allow for different time scales in the evolution
of $x$ and $y$ variables. 
To test the performance of the reservoir in the case of the nonlinear oscillations,
we set the FHN system to operate in the nonlinear oscillatory regime delimited in $\upalpha \in(-1,1)$
(in the present integrations $\upalpha =+0.5$ is used). 
$Q$ is an optional input current which can be 
 constant or can vary with time, depending on the
particular application. Because nonlinear FHN oscillations are possible even in the absence of input 
current, for simplicity we set $Q=0$. A typical ($x - y$) phase space plot of the limit cycle
oscillations is displayed in Fig.~\ref{fig01}(c). The system rapidly enters a closed limit cycle trajectory
and its fractal dimension is $d_f=1$, topologically identical to the classical circular trajectory
with a constant radius.  
\par All three models (Lorenz, R\"ossler and FHN) were integrated using the 4th order
Runge-Kutta method, with $dt=0.01$. Recording from the numerical integrations, we prepared 
the training and testing data sets
as time series consisting of 50,000 data points representing the $x$-component, 
for each model. The first $K_{\rm transient}= 8,000$ data points were
disregarded as transient data, $K_{\rm training}=10,000$ or 30,000 data points were used as training data, and 
$K_{\rm test}=400$ points were used as test data for predictions in all cases (see also Appendix~\ref{appendixB}).

\section{Results \& Predictions}
\label{sec:Results}
\subsection{Predictions of the Erd\"os-R\'enyi Reservoir}
\label{sec:predicitions-ER}

In this section, we test the predictions of the ER reservoir using the three representative 
nonlinear dynamical systems discussed in the previous section.   
\par Starting with ER predictions on the Lorenz model, 
we use a training data set containing
$K_{\rm training}$=30,000 training points and all other parameters 
as in the working set described
in Sec.~\ref{sec:Lorenz}.  For comparison, 
results using a smaller training set, with $K_{\rm training}$=10,000 data points, are provided in Appendix~\ref{appendixC}.
In Fig.~\ref{fig02} we present the reservoir performance on the Lorenz system 
in a color-coded map
as a function  of the reservoir size $N$ and the ER reservoir connectivity $p$.
The parameters used in the OLT\_ESN algorithm are provided in the caption of the figure. 
Especially, low learning rates are
typically used for chaotic data and, in the present study, values
of the order $r_{\rm learning} \sim 1/N$ are used.
\par In particular, in Fig.~\ref{fig02}(a) we present
the color-coded map of the reservoir performance for relatively large values of $p$,
spanning the interval $\left[ 0.1 , 1.0 \right]$ while the $N$ values range 
 as small as 100 nodes up to 1000 nodes. For statistical reasons, in the map (Fig.~\ref{fig02}(a))
the NRMSE is averaged over 20 different realizations of the ER reservoir.
We note that the performance gets the best values 
in the lowest part of the map, where $p$ values are as small as $p\sim 0.1$.
  This aligns with empirical values resulting from extensive  
 numerical studies reporting that sparse reservoirs perform best, 
and typical reservoir connectivities
used in the literature are of the order of 0.01. 

\par To explore further
the reservoir performance in sparser networks,
we present in Fig.~\ref{fig02}(b) a more exhaustive search in the range $p \in \left[ 0.001, 0.01 \right]$, for a single network realization.
For these low $p$ values the performance is better than for $p \in \left[ 0.1, 1.0 \right]$, since the NRMSE values 
reach very low values mostly below 0.1,  as compared to the case of Fig.~\ref{fig02}(a), where the performance reaches considerably high values in the largest 
part of the parameter map.  
\begin{figure}
\includegraphics[clip,width=0.45\textwidth,angle=0]{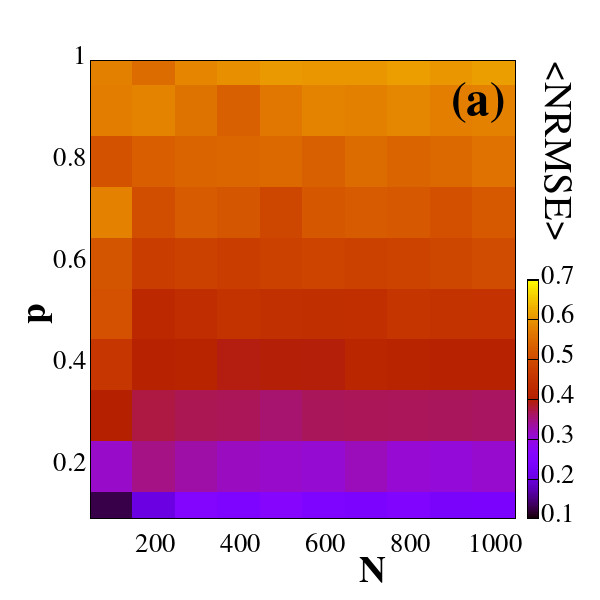}
\includegraphics[clip,width=0.45\textwidth,angle=0]{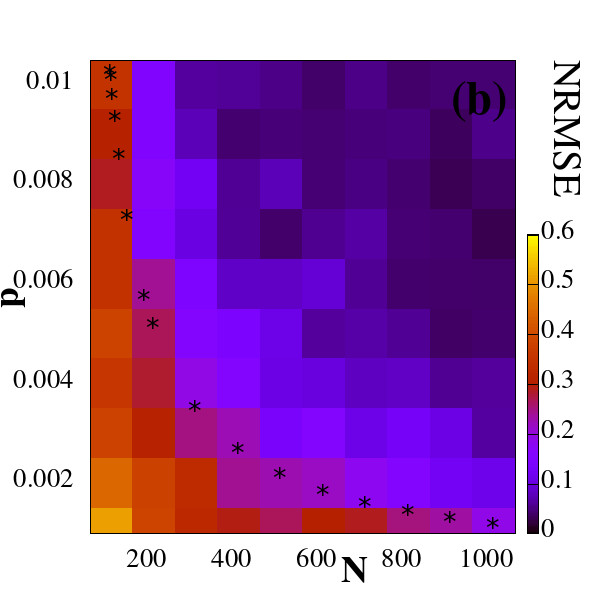}
\includegraphics[clip,width=0.45\textwidth,angle=0]{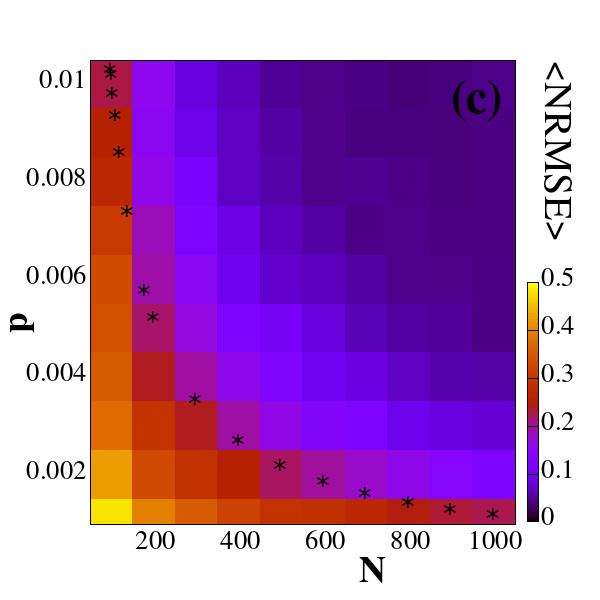}
\includegraphics[clip,width=0.45\textwidth,angle=0]{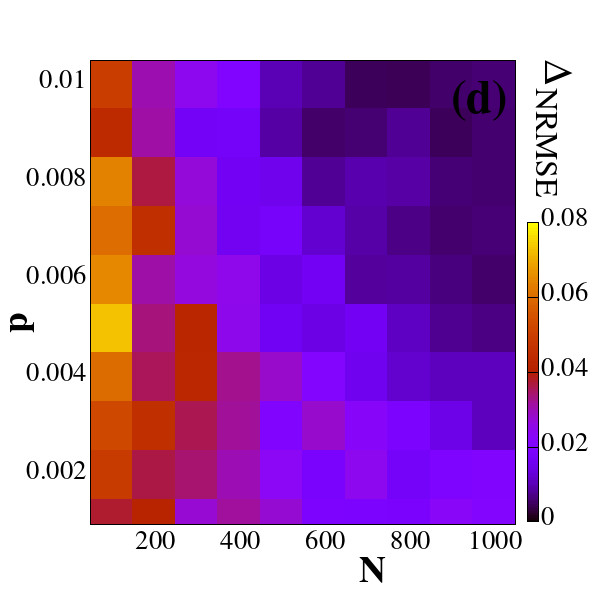}
\caption{\label{fig02} 
Lorenz chaotic dynamics:
 Color-coded map of the reservoir performance (NRMSE) 
as a function of the reservoir size $N$ and
the Erd\"os-R\'enyi reservoir connectivity $p$. (a) The case of large values of $p \in \left[ 0.1 , 1 \right] $ (averages over 20 realizations),
 (b) the case of smaller $p$ values (sparse reservoir), $p \in \left[ 0.001 , 0.01 \right] $, over one ER reservoir realization,
(c) the average reservoir performance, $\left< {\rm NRMSE} \right>$, over 20 realizations of the
ER network and (d) the performance deviation, $\Delta_{\rm NRMSE}$, over the 20 realizations.
In panels (b) and (c) the stars denote the critical point values of mean-field bond percolation. 
Other parameters are: $K_{\rm training}=30,000$, $b=1$,
$\alpha_{\rm leaky}=0.9$, $r_{\rm learning}=1/N$. 
In all cases, $K_{\rm transient} =8,000$ data points
were omitted as transient data. Parameters
of the Lorenz model data are provided in Sec.~\ref{sec:Lorenz}.
}
\end{figure}
\par Here, we would like to point out that this optimal reservoir performance may be related
to the critical point of percolation on the following grounds \cite{stauffer:1992}: 
In any network, the percolation
threshold, $p_c$, designates the lowest link density where the ``infinite cluster'' is formed.
The ``infinite cluster'' is defined as a continuous and connected pathway spanning
 across the entire network. 
At densities lower than $p_c$, the network (reservoir) is composed of many small isolated
clusters which do not ``communicate'' with each other. $p_c$ is the lowest 
link density where the large ``infinite cluster'' is formed, spanning the entire system
and thus allowing
intercommunication between all parts of the network. When the link density in the network exceeds
the percolation threshold, $p>p_c$, the ``infinite cluster'' becomes more extended and
the number of possible interconnection paths
 between different nodes increases accordingly. The reservoir may, therefore, be expected
to operate more 
efficient at densities on or slightly above the percolation threshold, where sufficient connectivity
allows information to propagate through the network while avoiding the flow of unnecessary, redundant 
exchanges that occur at high densities, $p>>p_c$.
\par From the theory of percolation, it is known that $p_c$ depends strongly on the structure of
the network \cite{stauffer:1992}. We are interested in bond (link) percolation and, for the case of a large
random network, mean-field bond percolation leads to $p_c=1/(N-1)$. This is because the critical 
point is achieved when each node connects with exactly one other node in the network. Since each node
can connect with $(N-1)$ other nodes, the probability to connect with one of them is $1/(N-1)$,
which is precisely equal to $p_c$ for the case of bond percolation in large 
random networks.\footnote{We note that for other networks the percolation threshold
may vary substantially depending on
their structure. For example, in bond percolation on 2D square lattice $p_c=0.5$, for 2D triangular 
lattice $p_c  \approx 0.3473$, for 3D cubic lattice: $p_c \approx 0.2488$ and for
the Bethe lattice (Cayley tree) $p_c = 1/(z-1)$ where $z$ is the coordination number,
and similarly for other networks \cite{stauffer:1992}.}

\par To check this argument, in Fig.~\ref{fig02}(b) we also plot in stars
the values of $p_c$ for a random ER network of size $N=100 \dots 1000$. We can observe
that the best performance is achieved for connectivities above the $p_c$ values. In fact, below the
 $p_c$ values the reservoir performance deteriorates considerably. This indicates that even though
sparse reservoir connectivity is efficient in general, 
going below the percolation threshold is not recommended, because
the performance is degraded
leading to worse predictions. This can be attributed to the
inhomogeneous information flow within the reservoir below $p_c$, 
where the different parts of the reservoir may operate independently.  

\par The results presented in Fig.~\ref{fig02}(b) have been computed using only one random ER
network realization (configuration).
To examine the diversity in reservoir performance due to different network details,
 we run averages over 20 different realizations of the  ER network in all $(N,p)$ combinations
and record the 
average performance  in map Fig.~\ref{fig02}(c). When the average is considered, 
the fluctuations decrease and the regions of best reservoir
performance become more evident, although the general tendencies remain the same in
one or over many different realizations. The amplitude of the performance 
fluctuations (deviation $\Delta_{\rm NRMSE}$) is plotted in Fig.~\ref{fig02}(d).  
Figures ~\ref{fig02}(b) and (c) indicate the same low density
region where the reservoir performance remains high and, in both cases, these regions are delimited
by the critical point of bond percolation (starred curve). Qualitatively similar conclusions
are drawn when a smaller training set is used, see Appendix~\ref{appendixC}.

\par
\begin{figure}[t]
\includegraphics[clip,width=0.48\textwidth,angle=0]{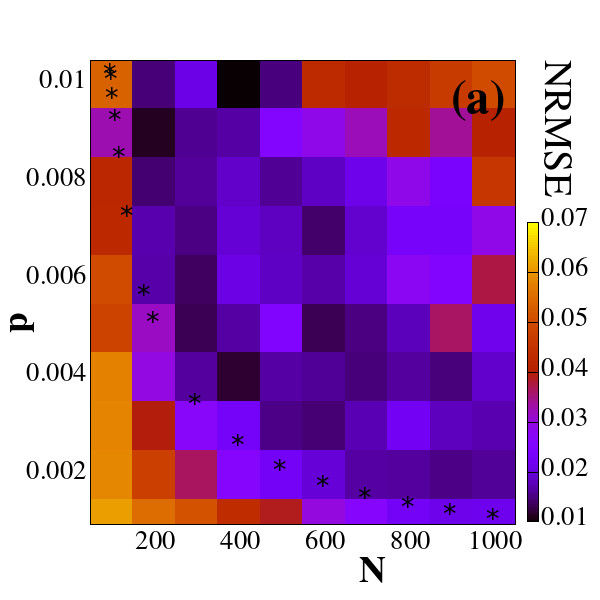}
\includegraphics[clip,width=0.48\textwidth,angle=0]{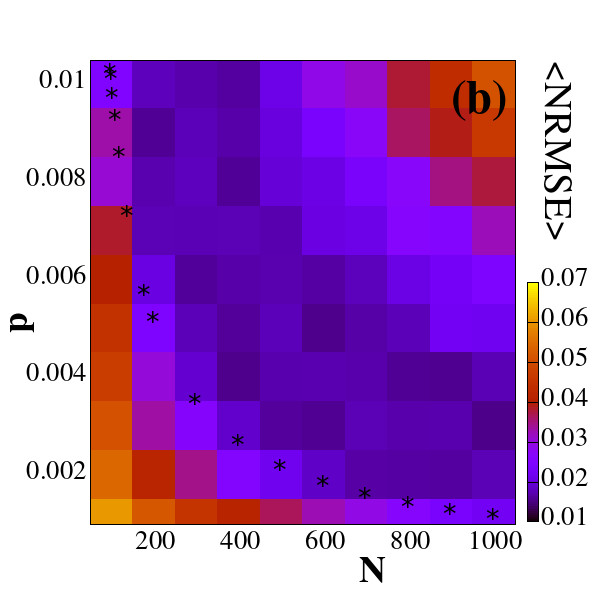}
\caption{\label{fig03} 
R\"ossler chaotic dynamics: (a) Color-coded map of the reservoir performance,
NRMSE, as a function of 
the reservoir size $N$ and the Erd\"os-R\'enyi reservoir connectivity $p$.  
Sparse reservoirs are used, with $p \in \left[ 0.001 , 0.01 \right] $.
Similarly to Fig.~\ref{fig02},
 the star points denote the critical point values of the mean-field bond percolation. 
(b) Average of the reservoir performance, $\left< {\rm NRMSE}\right>$, over 20 random network realizations.
Other parameters are: $K_{\rm training}=30,000$, $b=1$,
$\alpha_{\rm leaky}=0.9$, $r_{\rm learning}=1/(3N)$. In all cases $K_{\rm transient} =8,000$ data points
were omitted as transient data. The parameters
of the R\"ossler model data are provided in Sec.~\ref{sec:Roessler}.
}
\end{figure}

 We now turn our attention to the R\"ossler model in the chaotic parameter regions as was discussed
in Sec.~\ref{sec:Roessler}. We also use the ER reservoir
 to predict the next $K_{\rm test}=400$ points in the data series.
Relatively to the Lorenz model characterized by uniform evolution,
 the R\"ossler dynamics contains slow evolution periods
interrupted by relatively fast ones (complicated stretching-and-folding geometry). 
This multiscale character makes prediction more demanding. A typical way to overcome this complication
is to decrease the learning rate. For this reason, and to achieve reasonable prediction 
values (below 0.1), for
the predictions in the R\"ossler model we use lower learning rates, $r_{\rm learning}=1/(3N)$
(see details in Appendix~\ref{appendixA}).
To validate the reservoir  
performance we, first, plot the NRMSE
as a function of the variables $(N,p)$ for one ER network configuration, Fig.~\ref{fig03}(a). 
We also restrict the $p$ values in the range 
$\left[ 0.001 , 0.01\right] $ to compare with the percolation threshold of the ER network.
We note, that a band of $(N,p)$ values where the reservoir shows good performance appears just
after the bond percolation threshold of the ER network, depicted with stars in Fig.~\ref{fig03}(a). 
This conforms with the previous results on the Lorenz model following the idea that the 
ER reservoirs attain
high prediction performance when the connectivity spans through the entire network,
which is achieved on and beyond the bond percolation threshold.

\par As previously for the case of the Lorenz model, we also calculated the average performance
over 20 realizations of the ER network for the R\"ossler model data. 
The results are shown  in the map of Fig.~\ref{fig03}(b). The results for one network realization
(Fig.~\ref{fig03}(a)) and the average over the 20 network realizations (Fig.~\ref{fig03}(b)) do 
present significant qualitative similarities, with averaging primarily reducing 
realization-to-realization variability.
Again, optimal performances of the reservoir
are achieved in the region of the map that is delineated by the critical points of the 
mean-field bond percolation.

\par Regarding predictions on the FHN model,  in this case we test the ER reservoir predictions
when the FHN dynamics is in the limit cycle regime, as was discussed in Sec.~\ref{sec:FHN}. 
From Fig.~\ref{fig01}(c), we note that the limit cycle is not smooth but contains four abruptly
turning points. Before and after the turning points, fast and slow varying regimes are developed.
For this reason, we train the reservoir using the large FHN set  (30,000 points) as in the
case of the previous two models. The learning rate is kept to $r_{\rm learning}=1/N$ and
the rest of the reservoir hyperparameters remain the same as in the Lorenz and R\"ossler cases.
\begin{figure}[ht]
\includegraphics[clip,width=0.45\textwidth,angle=0]{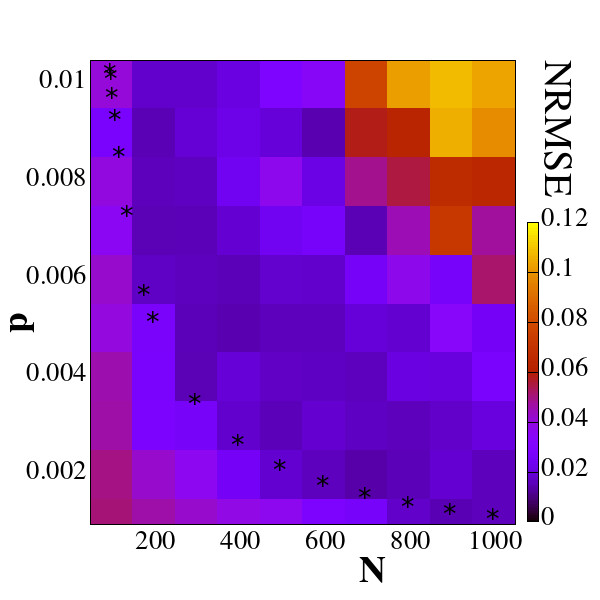}
\includegraphics[clip,width=0.45\textwidth,angle=0]{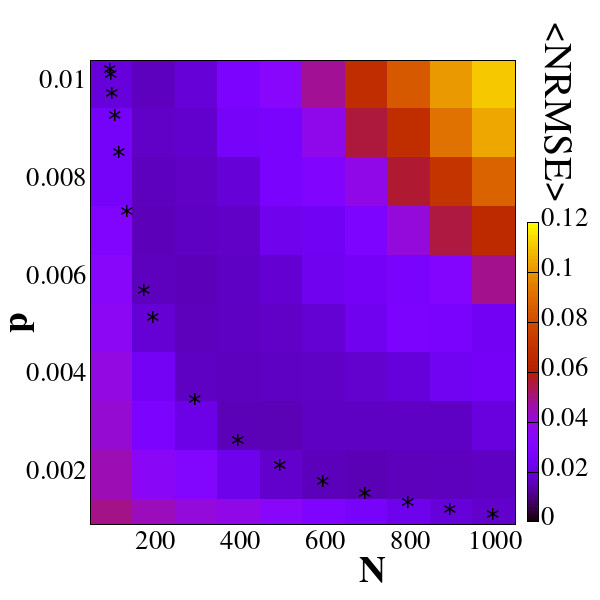}
\caption{\label{fig04} 
FitzHugh-Nagumo limit cycle dynamics: Color-coded maps of the reservoir performance (NRMSE) as a function of 
the reservoir size $N$ and the Erd\"os-R\'enyi reservoir connectivity $p$.  
Sparse reservoirs are used $p \in \left[ 0.001 , 0.01 \right] $.
The star points denote the critical point values of the mean-field bond percolation.
(a) The NRMSE values using one ER network realization, and (b) Average $\left< {\rm NRMSE} \right>$ 
over 20 ER
network realizations.
Other parameters are: $K_{\rm training}=30,000$, $b=1$,
$\alpha_{\rm leaky}=0.9$, $r_{\rm learning}=1/N$. In all cases $K_{\rm transient} =8,000$ data points
were omitted as transient data. The parameters
of the FHN model data are provided in Sec.~\ref{sec:FHN}.
}
\end{figure}

\par In Fig.~\ref{fig04}, the
performance as a function of the variables $(N,p)$ is presented, for a single 
network realization in Fig.~\ref{fig04}(a) and over 20 realizations
in Fig.~\ref{fig04}(b). Both plots give qualitatively similar predictions,
with  the local statistical
fluctuations decreasing and the NRMSE variations
becoming more gradual when averages are considered.
As usual, the starred curves represent the bond percolation critical density, $p_c$.
In the FHN case, 
we observe that even reservoirs of connectivity smaller than the critical one
(to the left of the starred line) may perform rather well.
This means that fewer (than critical) nodes can equally well capture the FHN dynamics. 
This can be due to the limited phase space of the FHN system, which is two-dimensional,
making the use of many nodes obsolete. 
However, in general, the best performances are computed 
when the reservoir connectivity is on the critical bond percolation values and to its 
immediate right regions.
In all cases, the performance deteriorates as the 
network connectivity increases further beyond the percolation threshold region.
In Appendix~\ref{appendixC}, results on FHN predictions using 
10,000 training data points are presented for comparison. 
The results in the Appendix are qualitatively similar to the ones 
presented here (using the larger training
set); this confirms that the limitations of the bond percolation for the reservoir
predictive power are  independent
of the size of the training set.

\subsection{Predictions of the Watts \& Strogatz Reservoir}
\label{sec:predicitions-WS}

Next, we test the predictive power of the Watts \& Strogatz (WS) reservoir 
using the time series of the three nonlinear dynamical systems.   
Circular networks lack the property of global randomness which was so important in the
previous section of ER reservoirs and which seems to statistically mimic the
brain of mammals. However, 
recent experimental studies have isolated invertebrate circuits 
which are organized as ring-like
networks, usually when the circuits  represent cyclic variables (like direction, phase, or rhythmic timing). A specific example is the fruit fly, Drosophila Melanogaster,
 whose central complex contains a beautifully organized ring attractor network \cite{seeling:2015,green:2017}. Noisy circular rings are
also reported in insect brains \cite{pisokas:2020}, 
reminiscent of the random shortcut connections in the WS networks.
 
\par To test the predictions of the WS Echo State Network, in this section
we use reservoirs of medium size, $N=600$ nodes,  arranged on a $k$-ring network.
Each node is connected to $R=k/2$ nodes to its left and $R$ nodes to
 its right. Shortcuts are introduced on the $k$-ring  with probability $p_{\rm WS}$,
in the sense that each one of the connections is cut with probability $p_{\rm WS}$
and subsequently rewired to a random position on the ring.
The performance is recorded as a function of the WS network parameters, $(k,p_{\rm WS})$,
and the results are presented below for the cases of Lorenz, R\"ossler and FHN models.
The parameter ranges studied are $2 \le k \le 20$ (or equivalently $1 \le R \le 10$)
and $0 \le p_{\rm WS} \le 1$.
For $k=2$ the ring nodes are connected to their nearest neighbors only. For $p_{\rm WS}=0$
the connectivity is deterministic on the $k$-ring, while for $p_{\rm WS}=1$ all the connections
have been randomly cut and pasted and this corresponds to a full random 
(i.e., Erd\"os-R\'enyi) network.

\par In the current section dealing with WS reservoir predictions, we use a different pallete for
the color-coded parameter maps, to differentiate with the maps in the previous section dealing
with ER reservoir predictions.

\par
\begin{figure*}
\includegraphics[height=0.48\linewidth, angle =0.0]{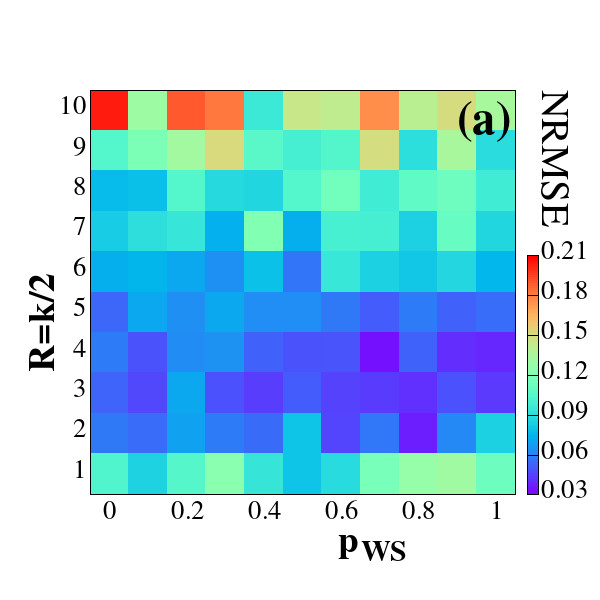}
\includegraphics[height=0.48\linewidth, angle =0.0]{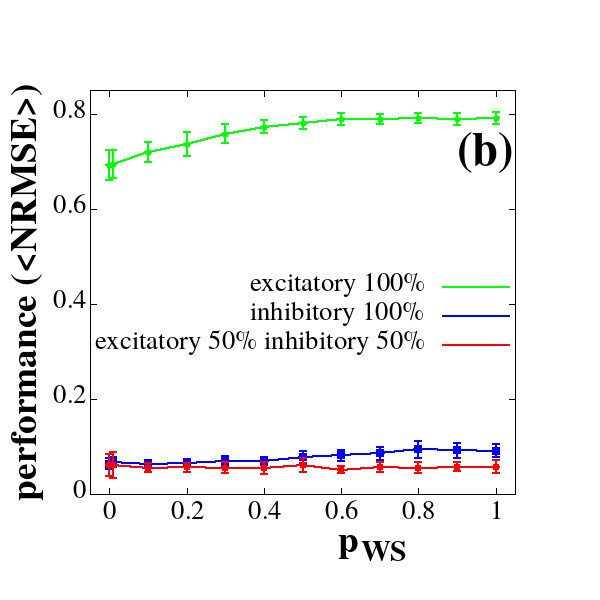}
\vspace{-1.5cm}
\includegraphics[height=0.48\linewidth, angle =0.0]{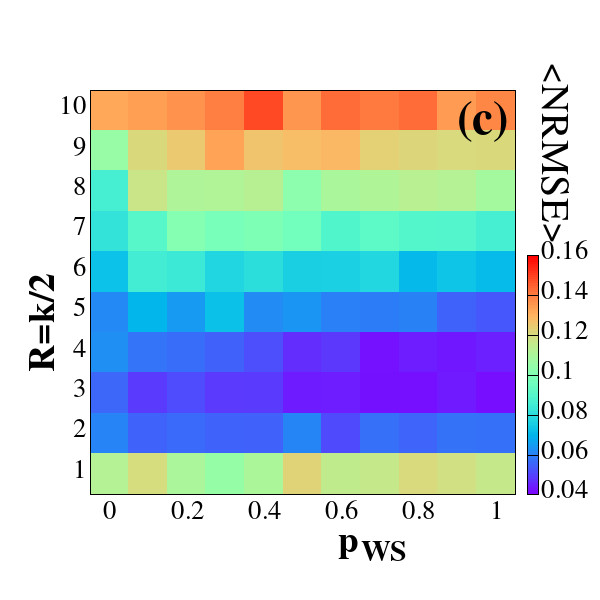}
\includegraphics[height=0.48\linewidth, angle =0.0]{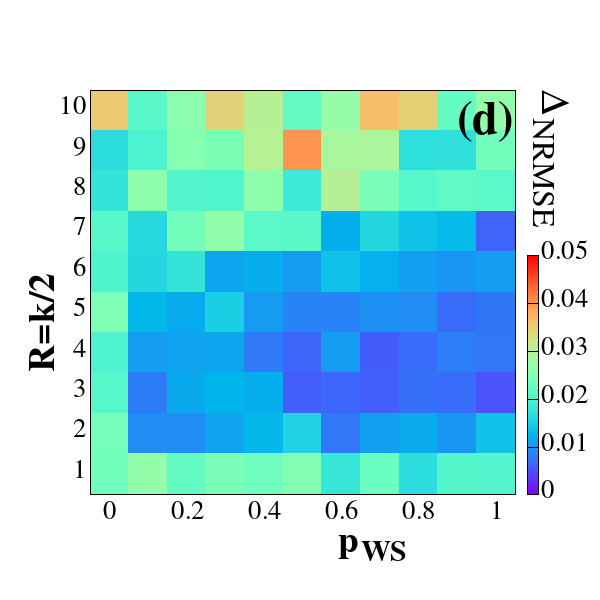}
\caption{\label{fig05}(Color online) 
Lorenz chaotic dynamics:
 Color-coded map of the reservoir performance (NRMSE) as a function of the
 Watts \& Strogatz rewiring probability $p_{\rm WS}$ and the
coupling range $R=k/2$.
 (a) The case of randomly distributed excitatory (positive)
and inhibitory (negative) coupling strengths, one network realization. 
 (b) For $R=2$, comparison of the average
reservoir performance when only positive coupling strengths
are considered (green line), only negative coupling strengths
(blue line) and when both positive and negative coupling strengths
are equally present (red line). (c) Average performance over 20 realizations
with randomly distributed excitatory 
and inhibitory couplings. 
(d) The performance deviation ($\Delta_{\rm NRMSE}$) over the 20 realizations.
Other parameters as in Fig.~\ref{fig02} and $N=600$. Parameters
of the Lorenz model data are provided in Sec.~\ref{sec:Lorenz}.
}
\end{figure*}

 We first present the predictive power of the WS reservoir on the Lorenz time series and
the results are shown in Fig.~\ref{fig05}. In Fig.~\ref{fig05}(a) we present the performance of the
WS reservoir when both inhibitory and excitatory links are present.  
The presence of inhibitory and excitatory connections is equiprobable in Fig.~\ref{fig05}(a), 
i.e., on average, half of the
connections are positive and the other half are negative. 
Overall, we note that the WS reservoir performance is best for 
low coupling ranges, $R=2,...5$,
and does not significantly depend on the rewiring probability $p_{\rm WS}$. 
This is in line with 
the idea that the network needs to be sparse. The case $R=1$ shows considerably inferior
performance. This can be attributed to the vulnerability
 of the nearest neighbor ring connectivity: If even one or two links are cut, the
ring may lose its percolation property, the network becomes disconnected
and the flow of links is interrupted.
For $R\ge 2$ it is less probable to create disconnected network domains even for large
rewiring probabilities.
  
\par As an additional case, in the WS reservoirs we also explore the significance 
of positive and negative coupling strengths and their influence on 
the reservoir prediction power.
In Fig.~\ref{fig05}(b) we present the 
average NRMSE in the case where 
only positive coupling weights are
present in the WS network (green line), when only negative couplings
are considered (blue line) and when 50\% positive and 50\% negative coupling weights
are considered (red line). 
The average is taken over 20 configurations of the reservoir weights. The error bars
indicating the standard deviations around the mean can be seen only in the case of
excitatory couplings, while in the case of inhibitory and mixed coupling they 
are smaller than, or of comparable size to the plotting symbols, 
and are therefore not clearly discernible.
As can be seen in the figure, the best performance is achieved
when both excitatory and inhibitory links are present in the system. The next best
performance is achieved for purely inhibitory links, while the worst performance is obtained
for purely excitatory links. These tests were performed for $R=2$, but similar conclusions
are drawn for the other values of $R$.

\par The above reported predictions are in line with previous studies showing that
the reservoirs perform better when both excitatory and inhibitory links are present
in the system. Recent investigations in biological neural networks agree that
in the human cerebral cortex the ratio of excitatory to inhibitory neurons is roughly 80\% to 20\%,
while locally the inhibitory connections can reach 30\% 
in many cases \cite{markram:2004,watts:2005,dehghani:2016,tatti:2017}.
Using ESNs for learning and prediction,
it would be interesting to investigate, in future works, which
ratio of excitatory to inhibitory connections gives the best predictions in ESNs and if this
ratio is consistent with the above biological observations in brain networks.

\par To test stochasticity issues on the WS reservoir predictions,
we average over 20 different realizations of the input
and the reservoir weights, while using the same WS network
configuration, see Fig.~\ref{fig05}(c). 
The prediction values are now confined further in the interval [0.04-0.16].
The average over many realizations reveals that 
the best predictions are obtained not only for small coupling ranges
but also they improve for large rewiring rates, $p_{\rm WS}$.
We remind that the large rewiring rates lead to random-like ER networks
and avoid the presence of isolated network domains.
In the parameter regions with short coupling
ranges $R$ and large rewiring probabilities $p_{\rm WS}$ we also
record the smallest fluctuations in the predictions, see Fig.~\ref{fig05}(d). 

\par To compare with the
Erd\"os-R\'enyi reservoir prediction on the Lorenz model in the previous section, 
 the average  connectivity of the WS network is
\begin{eqnarray}
p=\frac{2RN}{N(N-1)}=\frac{2R}{N-1},
\label{eq31}
\end{eqnarray}
because in the ring we have $N$ nodes and each one of them is connected to $2R$ nodes, i.e., we end with
 $2RN$ connections. Because the ring network can host $N(N-1)$ possible connections, the average connectivity is
then given by the ratio of these two factors, as in Eq.~\eqref{eq31}. Therefore, for the study in 
Fig.~\ref{fig05}(a) the connectivity  takes
values within the interval $0.003 < p < 0.03$ for $1 \le R \le 10$ and $N=600$ nodes.
For this range, the ER reservoir provides prediction power in the range ${\rm NRMSE}=
0.04-0.13$ for connectivities in the range $p=0.001-0.01$ (see from data
in Fig.~\ref{fig02}(c)). Approximately, in these ranges the WS model 
gives NRMSE$=0.04-0.14$ for $1 \le R \le 10$ and $N=600$ (see from data
in Fig.~\ref{fig05}(b)). The performance of the two reservoirs is then comparable in the prediction
 of the Lorenz chaotic data. However, if the coupling range is restricted to $2 \le R \le 5$,
the WS reservoir performs considerably better, with NRMSE in the range  $0.04-0.075$.

\par Regarding the aspect of percolation properties in the WS model, 
we note that in 1D (as is the k-ring network)
the bond percolation threshold is exactly calculated to be 1 for connections with the
nearest neighbors. This means that every node should be definitely connected with
its nearest neighbors, otherwise the system gets disconnected and information cannot cross
the network. Connections further than first neighbors can be considered as unnecessary
and often as burdensome, causing spreading of the information rather than directing it.
Therefore, the case $R=1$ and $p_{\rm WS}=0$
can allow optimal exchanges in the unrewired reservoir.
However, when rewiring is introduced in a ring with only nearest neighbors ($R=1$), 
the system may be split into disconnected clusters. This can be avoided when $R > 1$,
where multiple connections in each node guarantee the property of global connectivity,
even if  some of the local connections are rewired far away in the network.
 Indeed, for
the Lorenz model, the reservoir performs best for low coupling ranges, $2\le R \le 5$,
 both in the case of a
single realization (Fig.~\ref{fig05}(a)) and on average across realizations
(Fig.~\ref{fig05}(c)). In particular, both Figs.~\ref{fig05}(a) 
and (c) demonstrate that the performance
improves for $R$ values slightly higher than 1.
As explained above, this is because breaking and rewiring the 1D network
may cause isolations of different reservoir regions if $R=1$. For $R$ values slightly higher 
than 1, e.g. $R=2,3,4$, the isolated regions becomes less and less probable, the information
flows unobstructed through the reservoir and the NRMSE presents its lowest values.
Qualitatively similar results are obtained when using a smaller training set, see Appendix~\ref{appendixC}.

\par Next we discuss the predictions of the WS reservoir on the R\"ossler model  for different coupling ranges
and rewiring probabilities. We use the same WS coupling matrices as previously and
the same reservoir parameters, see caption of Fig.~\ref{fig05}. 
\begin{figure*}
\includegraphics[height=0.48\linewidth, angle =0.0]{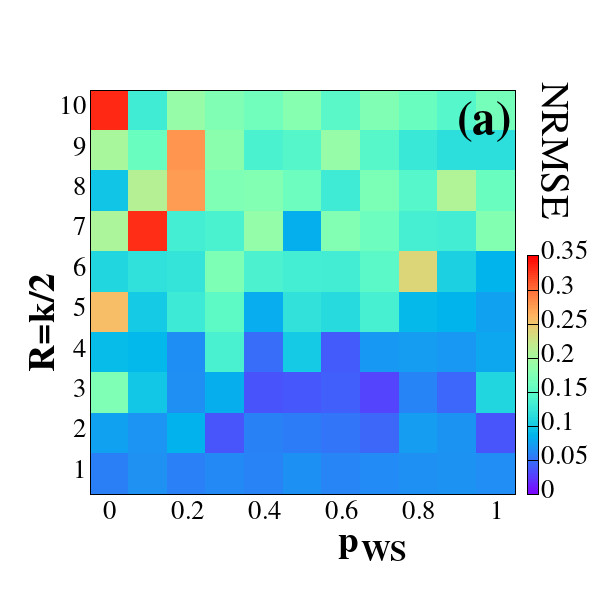}
\includegraphics[height=0.48\linewidth, angle =0.0]{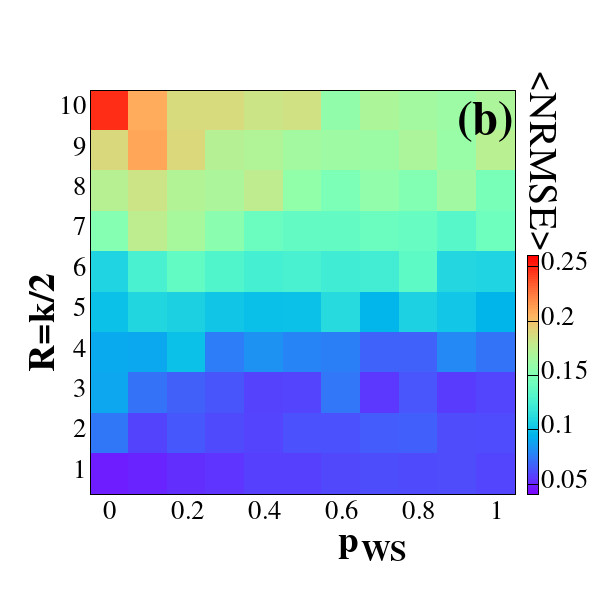}
\includegraphics[height=0.48\linewidth, angle =0.0]{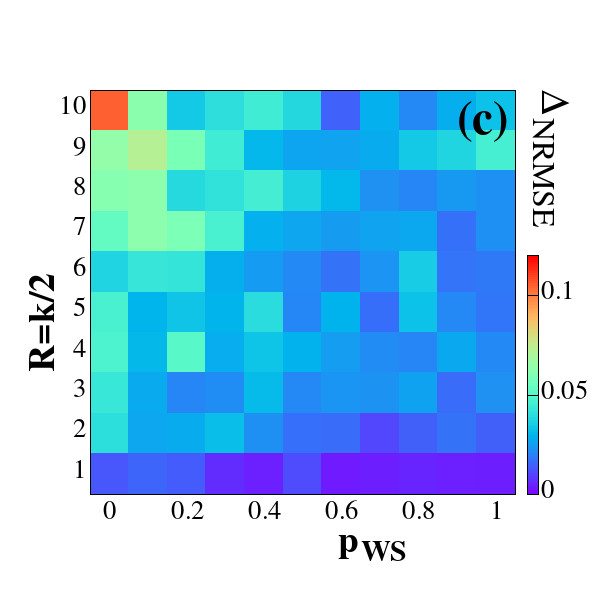}
\caption{\label{fig06}(Color online) 
R\"ossler chaotic dynamics:
 Color-coded map of the reservoir performance as a function of the
 Watts \& Strogatz rewiring probability $p_{\rm WS}$ and of the
coupling range $R=k/2$.
 In this case both excitatory (positive, 50\%)
and inhibitory (negative, 50\%) coupling strengths are randomly distributed on the $k-$ring. 
(a) Performance for one realization and (b) Average performance over 20 realizations.
Other parameters are: reservoir size $N=600$ nodes, $K_{\rm training}=30,000$, $b=1$,
$\alpha_{\rm leaky}=0.9$, $r_{\rm learning}=1/N$. 
In all cases $K_{\rm transient} =8,000$ data points
were omitted as transient data. The parameters
of the R\"ossler model data are provided in Sec.~\ref{sec:Roessler}.
}
\end{figure*}

\par The 
performance of the WS reservoir on predicting the chaotic R\"ossler model 
using one network realization are displayed in Fig.~\ref{fig06}(a),
where a color-coded map presents  the NRMSE performance of the reservoir
as a function of the coupling range $R$ and the rewiring probability $p_{\rm WS}$.
As a general conclusion, the performance range, interval $\left[0.0 , 0.35\right]$,
 is lower than in the Lorenz case
shown in Fig.~\ref{fig05}. In general, the R\"ossler model is harder to  learn and
predict due to the nonuniform dynamics consisting of interchanges of fast and slow
evolution.
Averaging over 20 configurations of input  and reservoir weights, slightly shrinks 
the prediction interval to $\left[ 0.04, 0.26\right]$. Here, best predictions are obtained 
for lower $R=1,2,3$ values. 

\begin{figure*} [hb]
\includegraphics[height=0.48\linewidth, angle =0.0]{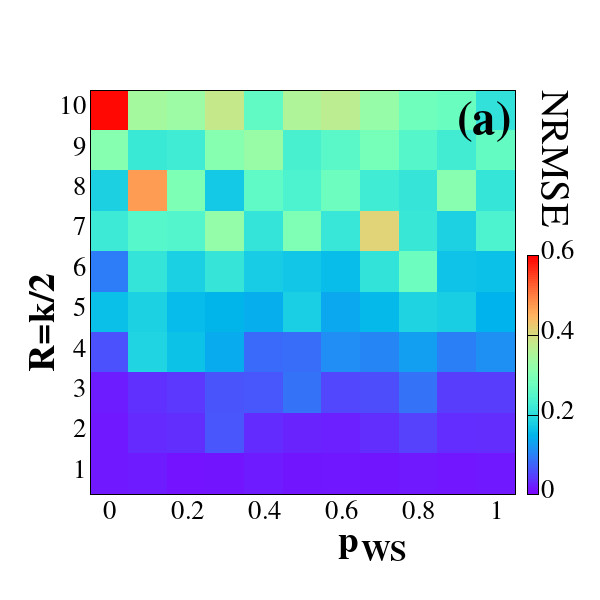}
\includegraphics[height=0.48\linewidth, angle =0.0]{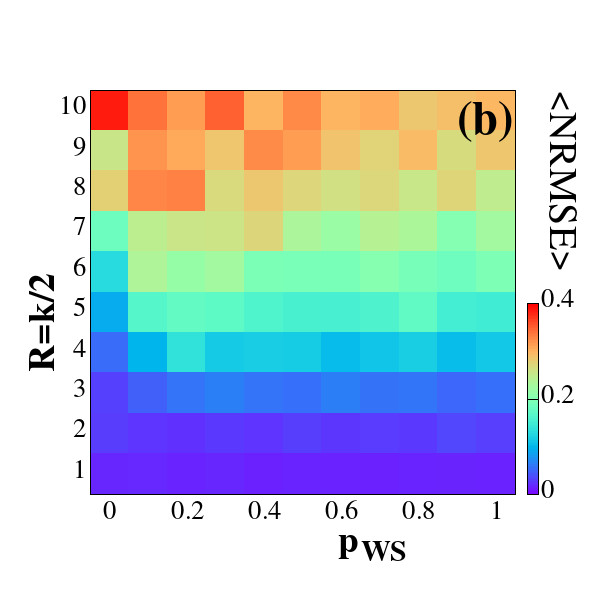}
\caption{\label{fig07}(Color online) 
FitzHugh-Nagumo limit cycle dynamics:
 Color-coded map of the reservoir performance (NRMSE) as a function of the
 Watts \& Strogatz rewiring probability $p_{\rm WS}$ and the
coupling range $R=k/2$.
Both excitatory (positive, 50\%)
and inhibitory (negative, 50\%) coupling strengths are randomly distributed on the $k-$ring. 
(a) Performance for one realization, (b) average performance over 20 realizations.
Other parameters are: reservoir size $N=600$ nodes, $K_{\rm training}=30,000$, $b=1$,
$\alpha_{\rm leaky}=0.9$, $r_{\rm learning}=1/N$. 
In all cases $K_{\rm transient} =8,000$ data points
were omitted as transient data. The parameters
of the FHN model data are provided in Sec.~\ref{sec:FHN}.
}
\end{figure*}

\par Regarding the WS reservoir predictions on the FHN model, in Fig.~\ref{fig07}(a),
we plot the network performance for a single realization of the WS network and
random and uniform choice of the coupling weights of the reservoir, as usual. Similar to
the cases of the Lorenz and R\"ossler models, low coupling ranges
produce better predictions than the higher ones. The same conclusions are 
reached if we average over 20 realizations of the input and reservoir weight
strengths, see Fig.~\ref{fig07}(b).
This means that we need only few neighborhood connections
 in order to produce reliable predictions, while the rewiring does not seem
to significantly influence the predictions.
We stress here that for the 20 different realizations  the same underlying WS network 
structure was retained and only the reservoir connecting weights 
and the input weights were randomly modified. 
The range of performances for the WS reservoir is $\left[ 0 - 0.4\right]$ for $1\le R \le 10$,
while for smaller coupling ranges $1 \le R \le 5$ the NRMSE drops in the interval $\left[ 0.01 - 0.12\right]$
The standard deviations of the performance over the 20 different realizations
is of the order 0.002-0.005 without any specific motif.

\par Overall, the WS reservoir produces comparable prediction rates 
 as the ER reservoir of the same size, with best
predictions provided by WS reservoirs at small coupling ranges $R$.
In the present study
statistics were taken into account with respect to the input and reservoir weight
values but not with respect to the WS reservoir connectivity architectures. This means that all the
reservoir connections were preserved in the considered configurations and only the connection
weights changed randomly from one configuration to another. In future studies, we
will need to extend the study using realizations where both the weights and the positions
of the couplings will be randomly selected. Averages over different link positions
may influence the predictive power of the WS reservoirs.

\section{Conclusions}
\label{sec:conclusion}
In the current study we address the ability of Online Trained Echo State Networks to
predict the trajectory of dynamical systems with chaotic (Lorenz and R\"ossler attractors)
and limit cycle dynamics (FitzHugh-Nagumo model). We first use a reservoir with Erd\"os-R\'enyi
connectivity and test its predictive power as a function of the reservoir size $N$ and the
 probability $p$ that any two nodes of the reservoir are connected. Our numerical investigations
show that the ER reservoir performs best in the $\left( N, p \right)$ region which 
is delineated by the bond
percolation curve. This region is characterized by small and sparse networks.  
In the case of Watts \& Strogatz reservoirs, the predictions
are in similar value intervals as in the ER ones. For all tested data
 the best performances are achieved when the coupling ranges are small,
close to nearest neighbor connectivity, while the rewiring probability does not seem
to affect significantly the WS predictive  power.
 In most of the above numerical investigations, both  positive and negative
coupling strengths were considered with relative ratio 1:1.

\par In view of these results, a number of open problems and extensions are proposed below for 
future studies.
\par In the case of the WS reservoir,
averages were taken over 20 different realization of the input and reservoir weight
values but not on different reservoir connection configurations. 
In future studies, the study must be extended to include realizations 
where both the weights and the rewired positions
 are randomly selected. Different structural network realizations
may influence the predictive power of the WS networks.

\par As earlier discussed, implementation of the Barabasi-Albert reservoir 
and other scale-free reservoir architectures could provide evidence on the
influence of long-range connectivity and on the presence of hubs in the reservoir performance.

\par Along with the presented computations, we have performed test runs for 
the case of symmetric reservoirs, where $w_{ij}=w_{ji}$, $i,j=1, \cdots N$,
and found that they achieve performances much worse than the purely random case
(results not shown).
Extending these ideas, other structure reservoir properties, such as spatial
reservoir correlations or networks with specific characteristics,
 need to be accessed in relation to performance.

\par In Sec.~\ref{sec:predicitions-WS}, the reservoir performance for the Lorenz model 
was studied and the best performance was achieved
when both excitatory and inhibitory links are  equally present in the system,
as opposed to pure inhibitory or pure excitatory connections. 
It would be interesting to investigate other ratios 
of excitatory to inhibitory connections in online (and offline) ESNs
and to check whether best performances are achieved when the excitatory-to-inhibitory
ratios take comparable values with the ones recorded in  
biological observations in brain networks
\cite{markram:2004,watts:2005,tatti:2017}.

\par In the present study we have only varied
the network connectivity parameters ($p$ and $N$  in the case of ER reservoir and
$R$ and $p_{\rm WS}$ in the case of the WS reservoir) in order to access the 
influence of the reservoir structural properties on its performance, comparatively. 
We have not scanned all
the reservoir parameters ($N$, $K_{\rm training}$, $b$,
$\alpha_{\rm leaky}$, $r_{\rm learning}$) in order to select
the ones that provide the absolute best performance for each network case.
One related question is whether the restriction of connectivity
on or around the bond percolation threshold holds for all values of the training
and testing parameters or only for restricted parameter regions.
In future
studies, scanning the reservoir structural parameters together with the training
and learning parameters will be necessary for selecting the parameter ranges
which produce the absolute best predictions.

\par Another direction worth following is to test different activation functions
such as ReLU or others, instead of the $\tanh$,  and compare whether they
improve predictability, or speed of computations. 
Instead of ESNs with online training 
 we may use typical ESNs
with ridge regression and compare predictive power, algorithmic speed and
other predictive properties.

\par  Typical physical reservoirs may also be tested, where the
nodal dynamics are extracted from a dynamical system. Of particular interest could be
dynamical systems representing neurons, such as leaky integrate-and-fire (LIF), FHN,
 HindMarsh-Rose models etc, and/or their equivalent
circuits. Do such reservoirs result in better predictions? Are there other advantages
in these alternative methods? Questions also related to the stability, complexity 
and speed of convergence of the
algorithms may be investigated along these research lines.

\begin{acknowledgments}
  This work was supported with computation time granted by the Greek Research \& Technology Network (GRNET) 
in the National High Performance Computing HPC facility -ARIS- under Project ID PR016032.   
\end{acknowledgments}

\bibliographystyle{iopart-num}
\bibliography{./provata2026.bib}

@inproceedings{jaeger:2001,
  author =        {Herbert Jaeger},
  booktitle =     {GMD report no. 148. GMD — German National Research
                   Institute for Computer Science.},
  title =         {The “echo state” approach to analysing and
                   training recurrent neural networks.},
  year =          {2001},
  url =           {http://www.faculty.iu-bremen.de/hjaeger/pubs/
                  EchoStatesTechRep.pdf},
}

@inproceedings{jaeger:2002,
  author =        {Herbert Jaeger},
  booktitle =     {Advances in Neural Information Processing Systems},
  editor =        {S. Becker and S. Thrun and K. Obermayer (Eds)},
  pages =         {593 -- 600},
  publisher =     {MIT Press},
  title =         {Adaptive Nonlinear System Identification with Echo
                   State Networks},
  volume =        {15},
  year =          {2002},
  url =           {https://proceedings.neurips.cc/paper_files/paper/2002/file/
                  426f990b332ef8193a61cc90516c1245-Paper.pdf},
}

@article{jaeger:2007,
  author =        {Herbert Jaeger and Mantas Lukosevicius and
                   Dan Popovici and Udo Siewert},
  journal =       {Neural Networks},
  pages =         {335 -- 352},
  title =         {Optimization and applications of echo state networks
                   with leaky-integrator neurons},
  volume =        {20},
  year =          {2007},
  doi =           {https://doi.org/10.1016/j.neunet.2007.04.016},
  url =           {},
}

@article{maass:2002,
  author =        {Maass, Wolfgang and Natschl\"ager, Thomas and
                   Markram, Henry},
  journal =       {Neural Computation},
  pages =         {2531 -- 2560},
  title =         {Real-time computing without stable states: a new
                   framework for neural computation based on
                   perturbations},
  volume =        {14},
  year =          {2002},
  doi =           {https://doi.org/10.1162/089976602760407955},
  url =           {},
}

@article{racca:2021,
  author =        {Racca, Alberto and Magri, Luca},
  journal =       {Neural Networks},
  pages =         {252 -- 268},
  title =         {Robust Optimization and Validation of Echo State
                   Networks for learning chaotic dynamics},
  volume =        {142},
  year =          {2021},
  doi =           {https://doi.org/10.1016/j.neunet.2021.05.004},
  url =           {https://www.sciencedirect.com/science/article/pii/
                  S0893608021001969},
}

@article{trierweiler:2021,
  author =        {Trierweiler Ribeiro, Gabriel and
                   Alves Portela Santos, Andr{\'{e}} and
                   Cocco Mariani, Viviana and
                   dos Santos Coelho, Leandro},
  journal =       {Expert Systems with Applications},
  pages =         {115490},
  title =         {Novel hybrid model based on echo state neural network
                   applied to the prediction of stock price return
                   volatility},
  volume =        {184},
  year =          {2021},
  doi =           {https://doi.org/10.1016/j.eswa.2021.115490},
  url =           {https://www.sciencedirect.com/science/article/pii/
                  S0957417421009003},
}

@article{heyder:2022,
  author =        {Heyder, Florian and Mellado, Juan Pedro and
                   Schumacher, J{\"o}rg},
  journal =       {Physical Review E},
  pages =         {055303},
  title =         {Generalizability of reservoir computing for
                   flux-driven two-dimensional convection},
  volume =        {106},
  year =          {2022},
  doi =           {https://doi.org/10.1103/PhysRevE.106.055303},
  url =           {},
}

@article{gonzalez:2022,
  author =        {Gonz{\'a}lez-Zapata, Astrid Maritza and
                   Tlelo-Cuautle, Esteban and Cruz-Vega, Israel},
  journal =       {Electronics},
  pages =         {3612},
  title =         {On the Optimization of Machine Learning Techniques
                   for Chaotic Time Series Prediction},
  volume =        {11},
  year =          {2022},
  url =           {https://www.mdpi.com/2079-9292/11/21/3612},
}

@article{sharifi:2023,
  author =        {Sharifi Ghazijahani, M. and Heyder, F. and
                   Schumacher, J. and Cierpka, C.},
  journal =       {Physics of Fluids},
  pages =         {115141},
  title =         {Spatial prediction of the turbulent unsteady von
                   Kármán vortex street using echo state networks Open
                   Access},
  volume =        {35},
  year =          {2023},
  doi =           {https://doi.org/10.1063/5.0172722},
}

@article{wang:2025,
  author =        {Wang, Xubin and Ma, Pei and Lian, Jing and
                   Liu, Jizhao and Ma, Yide},
  journal =       {Frontiers in Physics},
  title =         {An echo state network based on enhanced intersecting
                   cortical model for discrete chaotic system
                   prediction},
  volume =        {13 - 2025},
  year =          {2025},
  url =           {https://www.frontiersin.org/journals/physics/articles/
                  10.3389/fphy.2025.1636357},
}

@article{mandal:2025,
  author =        {Mandal, Pinak and Gottwald, Georg A.},
  journal =       {Nature Communications},
  pages =         {5961},
  title =         {Learning dynamical systems with hit-and-run random
                   feature maps},
  volume =        {16},
  year =          {2025},
  doi =           {https://doi.org/10.1038/s41467-025-61195-1},
  url =           {},
}

@article{yadav:2026,
  author =        {Yadav, Alok and Pasupa, Kitsuchart and
                   Loo, Chu Kiong},
  journal =       {IEEE Access},
  pages =         {124684 -- 124702},
  title =         {Evolutionary Echo State Network Architecture Search
                   for Time-Series Prediction},
  volume =        {14},
  year =          {2026},
  doi =           {10.1109/ACCESS.2026.3722373},
}

@article{pathak:2018,
  author =        {Pathak, Jaideep and Hunt, Brian and Girvan, Michelle and
                   Lu, Zhixin and Ott, Edward},
  journal =       {Physical Review Letters},
  pages =         {024102},
  title =         {Model-Free Prediction of Large Spatiotemporally
                   Chaotic Systems from Data: A Reservoir Computing
                   Approach},
  volume =        {120},
  year =          {2018},
  doi =           {10.1103/PhysRevLett.120.024102},
  url =           {https://link.aps.org/doi/10.1103/PhysRevLett.120.024102},
}

@article{lu:2018,
  author =        {Lu, Zhixin and Hunt, Brian R. and Ott, Edward},
  journal =       {Chaos},
  pages =         {061104},
  title =         {Attractor reconstruction by machine learning},
  volume =        {28},
  year =          {2018},
  url =           {https://doi.org/10.1063/1.5039508},
}

@article{vlachas:2020,
  author =        {P.R. Vlachas and J. Pathak and B.R. Hunt and
                   T.P. Sapsis and M. Girvan and E. Ott and
                   P. Koumoutsakos},
  journal =       {Neural Networks},
  pages =         {191 -- 217},
  title =         {Backpropagation algorithms and Reservoir Computing in
                   Recurrent Neural Networks for the forecasting of
                   complex spatiotemporal dynamics},
  volume =        {126},
  year =          {2020},
  url =           {https://doi.org/10.1016/j.neunet.2020.02.016},
}

@article{wikner:2024,
  author =        {Alexander Wikner and Joseph Harvey and
                   Michelle Girvan and Brian R. Hunt and
                   Andrew Pomerance and Thomas Antonsen and Edward Ott},
  journal =       {Neural Networks},
  pages =         {94 -- 110},
  title =         {Stabilizing machine learning prediction of dynamics:
                   Novel noise-inspired regularization tested with
                   reservoir computing},
  volume =        {170},
  year =          {2024},
  url =           {https://doi.org/10.1016/j.neunet.2023.10.054},
}

@article{li:2013,
  author =        {W.~K.~Li and M.~J.~Hausknecht and P.~Stone and
                   M.~D.~Mauk},
  journal =       {Neural Networks},
  pages =         {95--102},
  title =         {Neural Networks Using a million cell simulation of
                   the cerebellum: Network Scaling and Task Generality},
  volume =        {47},
  year =          {2013},
}

@inproceedings{schumacher:2015,
  address =       {Cham},
  author =        {Schumacher, Johannes and Toutounji, Hazem and
                   Pipa, Gordon},
  booktitle =     {Artificial Neural Networks},
  editor =        {Koprinkova-Hristova, Petia and Mladenov, Valeri and
                   Kasabov, Nikola K.},
  pages =         {63 -- 90},
  publisher =     {Springer International Publishing},
  title =         {An Introduction to Delay-Coupled Reservoir Computing},
  year =          {2015},
  doi =           {https://doi.org/10.1007/978-3-319-09903-3_4},
}

@article{pandey:2020,
  author =        {Pandey, Sandeep and Schumacher, J\"org},
  journal =       {Physical Review Fluids},
  pages =         {113506},
  title =         {Reservoir computing model of two-dimensional
                   turbulent convection},
  volume =        {5},
  year =          {2020},
  doi =           {10.1103/PhysRevFluids.5.113506},
  url =           {https://link.aps.org/doi/10.1103/PhysRevFluids.5.113506},
}

@article{arun:2024,
  author =        {Arun, R. and Sathish Aravindh, M. and Lakshmanan, M.},
  journal =       {Physical Review E},
  pages =         {034204},
  title =         {Reservoir computing with logistic map},
  volume =        {110},
  year =          {2024},
  doi =           {https://doi.org/10.1103/PhysRevE.110.034204},
  url =           {},
}

@article{arun:2025,
  author =        {Arun, R. and Sathish Aravindh, M. and Lakshmanan, M.},
  journal =       {Physical Review E},
  pages =         {015307},
  title =         {Spin-torque nano-oscillators for reservoir computing
                   applied to time-series data},
  volume =        {112},
  year =          {2025},
  doi =           {https://doi.org/10.1103/f2f3-l3fc},
  url =           {},
}

@article{thorne:2025,
  author =        {Braden J. Thorne and D{\'e}bora C. Corr{\^e}a and
                   Ayham Zaitouny and Michael Small and
                   Thomas J\"ungling},
  journal =       {Chaos},
  pages =         {023136},
  title =         {Reservoir computing approaches to unsupervised
                   concept drift detection in dynamical systems},
  volume =        {35},
  year =          {2025},
  doi =           {https://doi.org/10.1016/j.biopsych.2016.09.017},
  url =           {},
}

@article{xu:2025,
  author =        {Xu, Yinhao and Gottwald, Georg A. and Kuncic, Zdenka},
  journal =       {APL Machine Learning},
  pages =         {046104},
  title =         {Learning chaotic dynamics with neuromorphic network
                   dynamics},
  volume =        {3},
  year =          {2025},
  doi =           {https://doi.org/10.1063/5.0285089},
}

@inproceedings{xu:2025_2,
  author =        {Xu, Yinhao and Gottwald, Georg A. and Kuncic, Zdenka},
  booktitle =     {2025 International Joint Conference on Neural
                   Networks (IJCNN)},
  number =        {},
  pages =         {1 -- 8},
  title =         {Dynamic Reservoir Computing with Physical
                   Neuromorphic Networks},
  volume =        {},
  year =          {2025},
  doi =           {10.1109/IJCNN64981.2025.11229247},
}

@article{amann:2026,
  author =        {Andreas Amann and Kathy L{\"u}dge and Ulrich Parlitz and
                   Michael Small},
  journal =       {Chaos},
  pages =         {060402},
  title =         {Nonlinear dynamics of reservoir computing: Theory,
                   realization, and application},
  volume =        {36},
  year =          {2026},
  doi =           {https://doi.org/10.1063/5.0342148},
}

@article{baxevanis:2026,
  author =        {Baxevanis, George and L{\"u}dge, Kathy and
                   Hizanidis, Johanne},
  journal =       {arXiv preprint arXiv:2605.13354},
  title =         {Reservoir Computing with a single Josephson junction},
  year =          {2026},
}

@article{rathor:2026,
  author =        {Rathor, Shailendra K. and Jaurigue, Lina and
                   Ziegler, Martin and Schumacher, J{\"o}rg},
  journal =       {Chaos},
  pages =         {033117},
  title =         {Prediction performance of random reservoirs with
                   different topology for nonlinear dynamical systems
                   with different number of degrees of freedom},
  volume =        {36},
  year =          {2026},
  doi =           {https://doi.org/10.1063/5.0314081},
  url =           {},
}

@article{erdos:1959,
  author =        {Paul Erd\"os and Alfr\'ed R\'enyi},
  journal =       {Publicationes Mathematicae Debrecen},
  pages =         {290 -- 297},
  title =         {On random graphs I},
  volume =        {6},
  year =          {1959},
  doi =           {https://doi.org/10.5486/PMD.1959.6.3-4.12},
}

@book{barabasi:2016,
  author =        {A.~L.~Barabasi},
  publisher =     {Cambridge University Press, Cambridge},
  title =         {Network Science},
  year =          {2016},
  doi =           {},
}

@article{watts:1998,
  author =        {Duncan J. Watts and Steven H. Strogatz},
  journal =       {Nature},
  pages =         {440 -- 442},
  title =         {Collective dynamics of 'small-world' networks},
  volume =        {393},
  year =          {1998},
  doi =           {https://doi.org/10.1038/30918},
}

@article{widrow:1960,
  author =        {Bernard Widrow and Marcian E. Hoff},
  journal =       {IRE WESCON Conv. Rec.,},
  pages =         {96 -- 104},
  title =         {Adaptive switching circuits},
  volume =        {4},
  year =          {1960},
  doi =           {http://doi.org/10.1109/PROC.1975.10036},
}

@article{koford:1966,
  author =        {Koford, J. and Groner, G.},
  journal =       {IEEE Transactions on Information Theory},
  pages =         {42 -- 50},
  title =         {The use of an adaptive threshold element to design a
                   linear optimal pattern classifier},
  volume =        {12},
  year =          {1966},
  doi =           {https://doi.org/10.1109/TIT.1966.1053856},
}

@book{newman:2010,
  author =        {M.~Newman},
  publisher =     {Oxford University Press, Oxford},
  title =         {Networks: An Introduction, 1st edition},
  year =          {2010},
  doi =
  {https://doi.org/10.1093/acprof:oso/9780199206650.001.0001},
}

@article{albert:2002,
  author =        {R.~Albert and A.~L.~Barabasi},
  journal =       {Reviews of Modern Physics},
  pages =         {47--97},
  title =         {Statistical Mechanics of Complex Networks},
  volume =        {74},
  year =          {2002},
}

@article{bassett:2006,
  author =        {Bassett, D. S. and Bullmore, E.},
  journal =       {Neuroscientist},
  pages =         {512-523},
  title =         {Small-world brain networks},
  volume =        {12},
  year =          {2006},
}

@article{muldoon:2016,
  author =        {Muldoon, Sarah Feldt and Bridgeford, Eric W. and
                   Bassett, Danielle S.},
  journal =       {Scientific Reports},
  pages =         {22057},
  title =         {Small-World Propensity and Weighted Brain Networks},
  volume =        {6},
  year =          {2016},
  doi =           {https://doi.org/10.1038/srep22057},
  url =           {},
}

@article{hilgetag:2016,
  author =        {Hilgetag, Claus C. and Goulas, Alexandros},
  journal =       {Brain Structure and Function},
  pages =         {2361 -- 2366},
  title =         {Is the brain really a small-world network?},
  volume =        {221},
  year =          {2016},
  doi =           {https://doi.org/10.1007/s00429-015-1035-6},
  url =           {},
}

@article{liao:2017,
  author =        {Liao, Xuhong and Vasilakos, Athanasios V. and
                   He, Yong},
  journal =       {Neuroscience \& Biobehavioral Reviews},
  pages =         {286 -- 300},
  title =         {Small-world human brain networks: Perspectives and
                   challenges},
  volume =        {77},
  year =          {2017},
  doi =           {https://doi.org/10.1016/j.neubiorev.2017.03.018},
  url =           {https://www.sciencedirect.com/science/article/pii/
                  S0149763416307849},
}

@article{han:2025,
  author =        {Han, Yan-Kun and Zhang, Hai-Jun and Chen, Yu-Jing and
                   Liu, Chang and Zhang, Yu-He and Zhang, Zhan-Jun and
                   Jing, Run-Ting and Guo, Li and Li, Da and
                   Chu, Wen-Yue and Wu, Wen-Jun and Zhang, Kan and
                   Cui, Long-Biao},
  journal =       {Brain‐X},
  pages =         {e70025},
  title =         {Small-world network and neuroscience},
  volume =        {3},
  year =          {2025},
  doi =           {https://doi.org/10.1002/brx2.70025},
  url =           {https://onlinelibrary.wiley.com/doi/abs/10.1002/brx2.70025},
}

@article{eguiluz:2005,
  author =        {Egu{\'{\i}}luz, Victor M. and Chialvo, Dante R. and
                   Cecchi, Guillermo A. and Baliki, Marwan and
                   Apkarian, A. Vania},
  journal =       {Physical Review Letters},
  pages =         {018102},
  title =         {Scale-Free Brain Functional Networks},
  volume =        {94},
  year =          {2005},
  doi =           {10.1103/PhysRevLett.94.018102},
  url =           {https://link.aps.org/doi/10.1103/PhysRevLett.94.018102},
}

@article{lee:2010,
  author =        {Lee, UnCheol and Oh, GabJin and Kim, Seunghwan and
                   Noh, GyuJung and Choi, ByungMoon and
                   Mashour, George A.},
  journal =       {Anesthesiology},
  pages =         {1081 -- 1091},
  title =         {Brain Networks Maintain a Scale-free Organization
                   across Consciousness, Anesthesia, and Recovery},
  volume =        {113},
  year =          {2010},
  doi =           {10.1097/ALN.0b013e3181f229b5},
  url =           {},
}

@article{expert:2011,
  author =        {P.~Expert and R.~Lambiotte and D.~R.~Chialvo and
                   K.~Christensen and H.~J.~Jensen and D.~J.~Sharp and
                   F. ~Turkheimer},
  journal =       {Journal of The Royal Society Interface},
  number =        {57},
  pages =         {472--479},
  title =         {Self-similar correlation function in brain
                   resting-state functional magnetic resonance imaging},
  volume =        {8},
  year =          {2011},
}

@article{ribeiro:2021,
  author =        {Ribeiro, Tiago L. and Chialvo, Dante R. and
                   Plenz, Dietmar},
  journal =       {Frontiers in Systems Neuroscience},
  title =         {Scale-Free Dynamics in Animal Groups and Brain
                   Networks},
  volume =        {14 -- 2020},
  year =          {2021},
  doi =           {10.3389/fnsys.2020.591210},
  url =           {https://www.frontiersin.org/journals/systems-neuroscience/
                  articles/10.3389/fnsys.2020.591210},
}

@article{jones:2023,
  author =        {Jones, Sabrina A and Barfield, Jacob H and
                   Norman, V Kindler and Shew, Woodrow L},
  editor =        {Berman, Gordon J and Behrens, Timothy E and
                   Hengen, Keith B},
  journal =       {eLife},
  pages =         {e79950},
  title =         {Scale-free behavioral dynamics directly linked with
                   scale-free cortical dynamics},
  volume =        {12},
  year =          {2023},
  doi =           {10.7554/eLife.79950},
  url =           {https://doi.org/10.7554/eLife.79950},
}

@article{lorenz:1963,
  author =        {Edward N. Lorenz},
  journal =       {Journal of the Atmospheric Sciences},
  pages =         {130 -- 141},
  title =         {Deterministic nonperiodic flow},
  volume =        {20},
  year =          {1963},
  doi =
  {https://doi.org/10.1175/1520-0469(1963)020<0130:DNF>2.0.CO;2},
}

@article{roessler:1976,
  author =        {Roessler, Otto E.},
  journal =       {Physics Letters A},
  pages =         {397 -- 398},
  title =         {An Equation for Continuous Chaos},
  volume =        {57},
  year =          {1976},
  doi =           {http://doi.org/10.1016/0375-9601(76)90101-8},
}

@article{fitzhugh:1961,
  author =        {R. ~FitzHugh},
  journal =       {Biophysical Journal},
  pages =         {445--466},
  title =         {Impulses and physiological states in theoretical
                   models of nerve membrane},
  volume =        {1},
  year =          {1961},
  doi =           {https://doi.org/10.1016/S0006-3495(61)86902-6},
}

@article{hodgkin:1952,
  author =        {A.~L.~Hodgkin and A.~F.~Huxley},
  journal =       {The Journal of Physiology},
  pages =         {500--544},
  title =         {A quantitative description of membrane current and
                   its application to conduction and excitation in the
                   nerve},
  volume =        {177},
  year =          {1952},
  doi =           {10.1113/jphysiol.1952.sp004764},
}

@article{nagumo:1962,
  author =        {J.~Nagumo and A.~Arimoto and S.~Yoshizawa},
  journal =       {Proc. Inst. Radio Engineers},
  pages =         {2061},
  title =         {An active pulse transmission line simulating nerve
                   axon},
  volume =        {50},
  year =          {1962},
}

@article{rontogiannis:2021,
  author =        {A.~Rontogiannis and A.~Provata},
  journal =       {European Physical Journal B},
  pages =         {97},
  title =         {Chimera states in {FitzHugh-Nagumo} networks with
                   reflecting connectivity},
  volume =        {94},
  year =          {2021},
  doi =           {https://doi.org/10.1140/epjb/s10051-021-00097-9},
}

@book{stauffer:1992,
  author =        {Stauffer, Dietrich and Aharony, Ammon},
  publisher =     {Taylor \and Francis},
  title =         {Introduction To Percolation Theory, Second Edition},
  year =          {1992},
  doi =           {},
}

@article{seeling:2015,
  author =        {Johannes D. Seelig and Vivek Jayaraman},
  journal =       {Nature},
  pages =         {186 -- 191},
  title =         {Neural dynamics for landmark orientation and angular
                   path integration},
  volume =        {521},
  year =          {2015},
  doi =           {https://doi.org/10.1038/nature14446},
}

@article{green:2017,
  author =        {Jonathan Green and Atsuko Adachi and Kunal K. Shah and
                   Jonathan D. Hirokawa and Pablo S. Magani and
                   Gaby Maimon},
  journal =       {Nature},
  pages =         {101 -- 106},
  title =         {A neural circuit architecture for angular integration
                   in Drosophila},
  volume =        {546},
  year =          {2017},
  doi =           {https://doi.org/10.1038/nature22343},
}

@article{pisokas:2020,
  author =        {Ioannis Pisokas and Stanley Heinze and Barbara Webb},
  journal =       {eLife},
  pages =         {1 -- 49},
  title =         {The head direction circuit of two insect species},
  volume =        {9},
  year =          {2020},
  doi =           {https://doi.org/10.7554/.53985},
}

@article{markram:2004,
  author =        {Henry Markram and Maria Toledo-Rodriguez and Yun Wang and
                   Anirudh Gupta and Gilad Silberberg and Caizhi Wu},
  journal =       {Nature Reviews Neuroscience},
  pages =         {793 -- 807},
  title =         {Interneurons of the neocortical inhibitory system},
  volume =        {5},
  year =          {2004},
  doi =           {https://doi.org/10.1038/nrn1519},
  url =           {},
}

@article{watts:2005,
  author =        {Jo Watts and Alex M. Thomson},
  journal =       {The journal of Physiology},
  pages =         {89 -- 97},
  title =         {Excitatory and inhibitory connections show
                   selectivity in the neocortex},
  volume =        {562},
  year =          {2005},
  doi =           {https://doi.org/10.1113/jphysiol.2004.076984},
  url =           {},
}

@article{dehghani:2016,
  author =        {Nima Dehghani and Adrien Peyrache and
                   Bartosz Telenczuk and Michel Le Van Quyen and
                   Eric Halgren and Sydney S. Cash and
                   Nicholas G. Hatsopoulos and Alain Destexhe},
  journal =       {Scientific Reports},
  pages =         {23176},
  title =         {Dynamic Balance of Excitation and Inhibition in Human
                   and Monkey Neocortex},
  volume =        {6},
  year =          {2016},
  doi =           {https://doi.org/10.1038/srep23176},
  url =           {},
}

@article{tatti:2017,
  author =        {Roberta Tatti and Melissa S. Haley and
                   Olivia K. Swanson and Tenzin Tselha and
                   Arianna Maffei},
  journal =       {Biological Psychiatry},
  pages =         {821 -- 831},
  title =         {Neurophysiology and regulation of the balance between
                   excitation and inhibition in neocortical circuits},
  volume =        {81},
  year =          {2017},
  doi =           {https://doi.org/10.1016/j.biopsych.2016.09.017},
  url =           {},
}

@article{widrow:1975,
  author =        {Bernard Widrow and Glover, J.R. and McCool, J.M. and
                   Kaunitz, J. and Williams, C.S. and Hearn, R.H. and
                   Zeidler, J.R. and Eugene Dong, Jr. and Goodlin, R.C.},
  journal =       {in Proceedings of the IEEE},
  pages =         {1692 -- 1716},
  title =         {Adaptive noise cancelling: Principles and
                   applications},
  volume =        {63},
  year =          {1975},
  doi =           {http://doi.org/10.1109/PROC.1975.10036},
}
\vskip 1cm

\section*{Appendix}
\vskip 0.3cm
\appendix
\section{Detailed OLT\_ESN algorithm}
\label{appendixA}
\par Before providing the detailed algorithm description, we recapitulate here all the
parameters and variables of the algorithm. 
\begin{itemize}
\item Parameter $N$. Represents the number of nodes of the reservoir.
\item Connectivity matrix ${\bf W}=\left[ w_{ij}\right]$. Has dimensionality $N \times N$ and its elements $w_{ij}$ represent the link weights between
pre- and post-synaptic nodes, $j$ and $i$, respectively. 
\item  Input weight vector, $\overrightarrow{W}^{\rm in}$. Has dimensionality $N$ and 
its elements $w^{\rm in}_i$ provide the weights that scale the data fed to nodes $i$,
 $i=1,2, \cdots N$. 
\item  Output weight vector, $\overrightarrow{W}^{\rm out}$. Has dimensionality $N+1$ and 
its elements $w^{\rm out}_i$ are adjusted every time a new incoming, training data point enters the reservoir.
\item Parameter $K_{\rm transient}$. Number of data points in the time series that are
disregarded as transient data.
\item Parameter $K_{\rm training}$. Size of the training data set.
\item Parameter $K_{\rm test}$. Size of the test set.
\item Data element $I^{\rm training}(n)$ (and $I^{\rm test}(n)$). The input
element of the training (and testing) sets fed to the reservoir at time step $n$.
\item Parameter $r_{\rm learning}$. It is a constant, small, learning rate to 
adjust the learning speed of the output weights.
\item Parameter $\alpha =\alpha_{\rm leaky}$. It is a leaky factor used to account 
for the percentage of past reservoir states brought forward at every training step.
\item Variable $n$. A counter of the points in the (training and testing) data sets. 
\item Vector variable $\overrightarrow X(n)$. Represents the state of the reservoir
at time $n$. The element $x_i(n)$ is the value of the node $i$ of the reservoir at
time step $n$, $i=1, \dots , N$.

\end{itemize}

\centerline{\bf{Initialization}}

\par The state of the reservoir at every time step $n=1, \dots , N_{\rm training}$ of the training period
is expressed as a vector $\overrightarrow X(n)$, where $x_i(n)$ is the value of the node $i$ 
of the reservoir at
time step $n$, $i=1, \dots , N$. We start by randomizing the input weights using typically 
a uniform distribution, $w^{\rm in}_i$, 
in the interval $(-1,1)$. Similarly, the elements of the reservoir connectivity matrix
$w_{ij}$ are also selected randomly in the interval $(-1,1)$,
 to allow for excitatory as well as
inhibitory interactions within the reservoir. 
The output weights $w^{\rm out}_i$, $i=1, \dots , N+1$,
are initially set to 0, but they change drastically during the
course of the training. We note that the output weights contain one more component than the
number of nodes. This bias, $b$, is added in order to allow the readout layer to 
take into account non-zero offsets of the data.
The initial values of the nodes of the reservoir $x_i(n=0)$ are also 
initialized to zero for all $i$.
To keep the algorithm as simple as possible, and for a better perception of the 
formation of the output
weights and of the prediction process in general, in the current implementation
we use 1D sequential data in the form of a time series. 

\centerline{\bf{The Training Stage}}

At every time step, 
$n=1, \dots , N_{\rm training}$,
of the Training Stage one
data point of value $I(n)$ is fed to all the nodes of the reservoir: 
\begin{enumerate}
\item The data value $I(n)$ is fed to all nodes scaled with the appropriate weights, $w^{\rm in}_i$, and
 the contribution $w^{\rm in}_i I(n)$ is calculated.
      \begin{eqnarray}
      C^{\rm input}_i(n+1)=w^{\rm in}_i I(n) .
      \label{eq01}
      \end{eqnarray}
\item In addition, the contribution to node $i$ due to the interactions with all other connected nodes $j$
       is calculated as
      \begin{eqnarray}
      C^{\rm inter}_i(n+1)=\sum_{j=1}^Nw_{ij}x_j(n) .
      \label{eq02}
      \end{eqnarray}
\item Then, the sum of $C^{\rm input}$ and $C^{\rm inter}$ is passed through an activation function. In the present study
       the hyperbolic tangent, $\tanh$, is used as the activation function and
      the value $x_i$ of each node $i$ at the next iteration step, $n+1$, is calculated as 
      \begin{eqnarray}
      x_i(n+1)=(1-\alpha)x_i(n)+\alpha \tanh \left[C^{\rm input}_i(n+1)+C^{\rm inter}_i(n+1)\right].
      \label{eq03}
      \end{eqnarray}
     In Eq.~\eqref{eq03}, $\alpha$ is the leaky parameter used to account for the percentage of past 
      reservoir states brought forward at 
      every training step. If $\alpha =1$ there is no memory in the reservoir states and
      they only get contributions from new input and other nodes at every training step.
\item Preparation of the augmented state. A new state $z_i$, $i=1, \dots , N+1$, is prepared as
      \begin{eqnarray}
      z_i=x_i, \>\>\> i=1, \dots , N,
      \nonumber \\
    z_{N+1}=b , \>\>\>\>\>\>\>\>\>
       \label{eq04}
      \end{eqnarray} 
      to account for potential offsets of the data and to add extra freedom to the reservoir. $b$ is
      called the bias and normally takes the value $b=1$.
\item Predict next data value. Using the output weights that have been computed in the
      previous training step the algorithm attempts to predict the next data value as
      \begin{eqnarray}
      I_{\rm predict}(n+1)=\sum_{i=1}^{N+1} w^{\rm out}_i(n) \> z_i .
      \label{eq05}
      \end{eqnarray}
\item Calculate the error between the predicted value $I_{\rm predict}(n+1)$ and the true
      data value $I(n+1)$ as:
             \begin{eqnarray}
      {\rm error}= I(n+1)-I_{\rm predict}(n+1).
      \label{eq06}
      \end{eqnarray}
\item Using the error from Eq.~\eqref{eq06}, the output weights get updated as:
             \begin{eqnarray}
      w^{\rm out}_i(n+1)= w^{\rm out}_i(n) -r_{\rm learning} \times {\rm error} \times z_i .
      \label{eq07}
      \end{eqnarray}
      Here, $r_{\rm learning}$ is a constant, small, learning rate to adjust the learning speed
      of the output weights. Using this adjustment, the output weight vector is transformed to hit
      in the direction of next data points. 
\item After all output weights are trained with influence of the $I(n)$ input data point,
     the algorithm returns to step 1, for a next data point of the training set to enter the reservoir.

\item After $N_{\rm training}$ time steps, all data points from the training set have entered the reservoir
      and the Training Stage is completed. 
The output weights are fixed and do not change
 in the next stages.
\end{enumerate}

\par The weights' update in step 7 follows the classical online 
learning and adaptive
filtering rule, also known as the ``delta rule'',
that was introduced by Bernard Widrow and Marcian Hoff in the 1960's. 
The key idea is the use of the
Least Mean Squares (LMS) adaptive algorithm which starts by assuming small weights 
(zero in most cases) 
and updates them at each step using the instantaneous gradient 
of the mean square error
(online adaptive).  This approach is also referred to as ``Widrow-Hoff LMS algorithm'',
 is nicely summarized in Appendix A of Ref.~\cite{widrow:1975} and is 
reported in Refs.~\cite{widrow:1960,koford:1966}.

\par It is also useful to discuss the reasons for augmenting the reservoir
state using the bias, $b$, in step 4. This extra bias node introduces additional freedom to the reservoir and
is most useful when the data is not centered around zero. Note that if at a certain time step the state
vector $\overrightarrow{X}=0$, then the predicted values in step 5 will be 0 and the output weights will not be
updated in step 7. The addition of even a small, constant bias creates an affine flexibility and the reservoir does not
get trapped in this null state. In special cases, the state vector can be expanded using additional features, 
such as nonlinear transformations ($x_i^2, x_i^3$),
 cross terms ($x_ix_j$), delayed states ($x_i(t-\tau )$) and
others. These types of augmented states are pertinent in different problems involving spatial or temporal data
correlations and will not be discussed in the present study. In general,
augmented states,  can push ESN performance closer to more complex models, without loosing the simplicity that makes reservoir computing attractive.

\par Regarding the learning rate parameter in Eq.~\eqref{eq07}, $r_{\rm learning}$, 
it needs to be adjusted with the reservoir
size to moderate influx of information leading occasionally to linearly dependent states and 
ultimately to very large output weights. 
Normally $r_{\rm learning} \sim 1/N$, and this scaling is used in the present study.

\centerline{\bf{The Testing Stage}}

\par After the Training Stage the algorithm enters into the Testing Stage. In this stage, the 
output weights are kept constant, as they were calculated at the end of the training stage.
These constant output weights are then used to predict a number of testing points, $N_{\rm test}$.
The predicted points are compared to the true data points and their deviations are used to evaluate 
the efficiency (performance, predictive power) of the ESN. The Testing Stage of the algorithm is
as follows. At every time step $n=1, \dots , N_{\rm test}$:
\begin{enumerate}
\item The data value $I^{\rm test}(n)$ from test data
      is fed to all nodes scaled with the appropriate weights, $w^{\rm in}_i$, and
      the contribution $w^{\rm in}_i I^{\rm test}(n)$ is calculated:
      \begin{eqnarray}
      C^{\rm input}_i(n+1)=w^{\rm in}_i I^{\rm test}(n) .
      \label{eq11}
      \end{eqnarray}
\item In addition, the contribution to node $i$ due to the interactions with all other 
       connected reservoir nodes $j$
       is calculated as
      \begin{eqnarray}
      C^{\rm inter}_i(n+1)=\sum_{j=1}^Nw_{ij}x_j(n) .
      \label{eq12}
      \end{eqnarray}
\item Then, the sum of $C^{\rm input}$ and $C^{\rm inter}$ is passed through the activation function $\tanh$,  and
      the value $x_i$ of each node $i$ is calculated as: 
      \begin{eqnarray}
      x_i(n+1)=(1-\alpha)x_i(n)+\alpha \tanh \left[C^{\rm input}_i(n+1)+C^{\rm inter}_i(n+1)\right] .
      \label{eq13}
      \end{eqnarray}
\item The new, augmented state $z_i$, $i=1, \dots , N+1$, is prepared as
      \begin{eqnarray}
      z_i=x_i, \>\>\> i=1, \dots ,  N \>\>\>,
      \nonumber \\
    z_{N+1}=b . \>\>\>\>\>\>\>\>\>
       \label{eq14}
      \end{eqnarray} 
\item Predict next test data value, $I^{\rm test}(n+1)$ . Using the constant output weights 
      that have been computed at the end of the Training Stage,
      the algorithm attempts to predict the next data value as
      \begin{eqnarray}
      I^{\rm test}_{\rm predict}(n+1)=\sum_{i=1}^{N+1} w^{\rm out}_i \> z_i .
      \label{eq15}
      \end{eqnarray}
\item Calculate the error between the predicted value $I^{\rm test}_{\rm predict}(n+1)$ and the true
      data value $I^{\rm test}(n+1)$ as:
             \begin{eqnarray}
      {\rm error}= I^{\rm test}(n+1)-I^{\rm test}_{\rm predict}(n+1).
      \label{eq16}
      \end{eqnarray}
\item Accumulate the performance index as the variance over all the errors produced 
      during the Testing Stages ( $N_{\rm test}$ test points). This is also known as Normalized Root Mean Square Error (NRMSE):
             \begin{eqnarray}
      {\rm NRMSE}= \frac{\sqrt{\sum_{n=1}^{N_{\rm test}}\left[ I^{\rm test}(n)-I^{\rm test}_{\rm predict}(n)  \right] ^2 }}{\sqrt{\sum_{n=1}^{N_{\rm test}}\left[ I^{\rm test}(n) \right] ^2}}. 
      \label{eq17}
      \end{eqnarray}
\item The algorithm returns to step 1, for a next test data point to enter the reservoir.

\item After $N_{\rm test}$ time steps, all test data points have entered the reservoir
      and the Testing Stage is completed. The value of the performance index, NRMSE in Eq.~\eqref{eq17},
 reflects the predictive power of  the reservoir.
\end{enumerate}

\eject
\section{Adjusting the size of the Test Set }
\label{appendixB}
\renewcommand{\thefigure}{B\arabic{figure}}
\setcounter{figure}{0}
\par
\begin{figure}[b]
\includegraphics[clip,width=0.6\textwidth,angle=0]{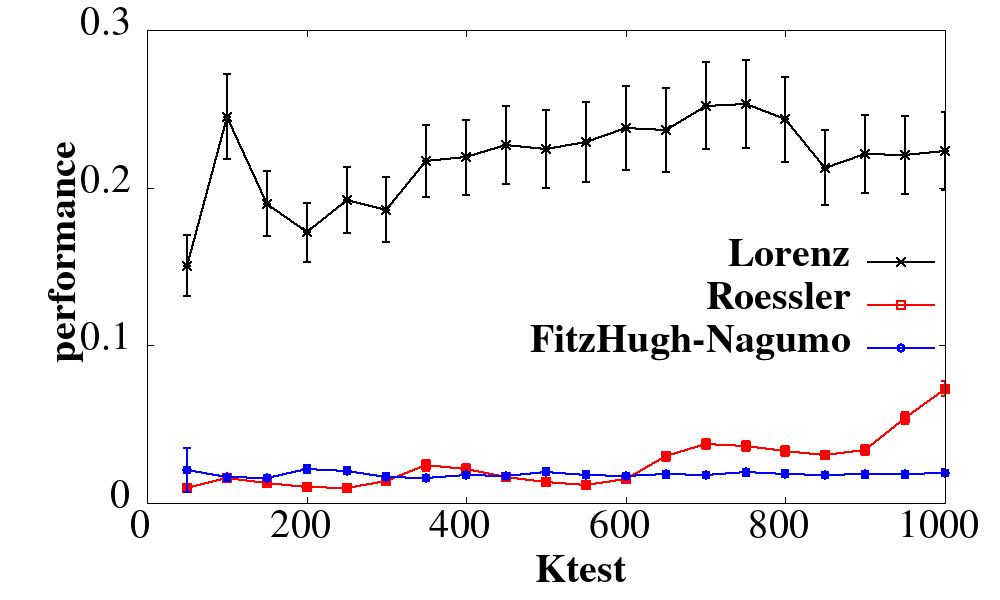}
\caption{\label{figB1} The scaling of the reservoir performance as a function
of the size of the testing set, $K_{\rm test}$. The black line represents the
prediction performance on the Lorenz data, the red line shows the
results on the R\"ossler data and the blue line the
results on the FHN data. Other parameters are: $N=500$, $p=0.002$, $K_{\rm training}=30,000$, $b=1$,
$\alpha_{\rm leaky}=0.9$, $r_{\rm learning}=1/N$ (1/3N for R\"ossler model). In all cases 8,000 data points
were omitted as transient data. The parameters used in the production 
of the Lorenz, R\"ossler and FHN data are described in Secs.~\ref{sec:Lorenz}, 
~\ref{sec:Roessler} and ~\ref{sec:FHN}, respectively.
}
\end{figure}

  To determine the appropriate 
 size of the test set, $K_{\rm test}$, we run the algorithm for different $K_{\rm test}$-values
to examine how the performance scales with the size of the test set. For these trial runs
we used a reservoir size of $N=500$ and 
reservoir connectivity of $p=0.002$. The results are plotted in Fig.~\ref{figB1} with other parameters
 provided in  the figure caption. 
\par The results in Fig.~\ref{figB1} show that the best performance is achieved for the FHN case,
which has the simplest attractor (limit cycle) amongst all three. Note that the error-bar sizes in the cases 
of R\"ossler and FHN attractors are of the order of the symbol sizes and are not well discernible in the image.
Regarding, in particular, the predictions of the R\"ossler model, a lower learning rate was used,
due to the different fast and slow timescales of the dynamics (see Sec.~\ref{sec:predicitions-ER}).
All predictions (including the Lorenz one) present a plateau at intermediate $K_{\rm test}$ values.
Based on these results, we select to work with $K_{\rm test}=400$ throughout this study, which, on the one hand, is
small enough to allow short run times and, on the other hand, produces reasonable performance outputs for
all three models.

\section{Effects of the size of the training set }
\label{appendixC}
\renewcommand{\thefigure}{C\arabic{figure}}
\setcounter{figure}{0}

{\bf{\it{- Erd\"os-R\'enyi reservoir predicting Lorenz dynamics}}}

\par
\begin{figure}[h]
\includegraphics[clip,width=0.6\textwidth,angle=0]{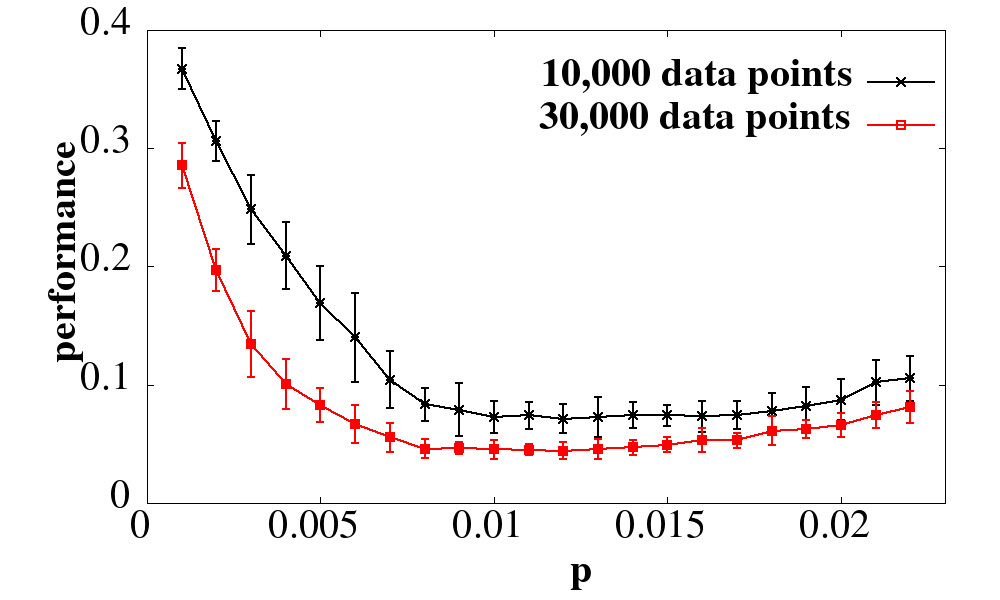}
\caption{\label{figC1} 
Comparison of predictions on the Lorenz model from an ER reservoir using a smaller and a 
larger training sets with $K_{\rm training}=10,000$ and  $30,000$, respectively.
All other parameters are as in Fig.~\ref{fig02}. 
}
\end{figure}

   To test the effects of the size of the training set, $K_{\rm training}$, we first present
a comparison between results on the ER predictions of the Lorenz dynamics using 
a smaller training data set containing 10,000 points as compared to the results in the main text,
where 30,000 data points were used. We observe, see Fig.~\ref{figC1}, that the larger
training set demonstrates better performance, while both training 
sets show a minimum NRMSE at the same $p$ region.
\par
\begin{figure}[]
\includegraphics[clip,width=0.45\textwidth,angle=0]{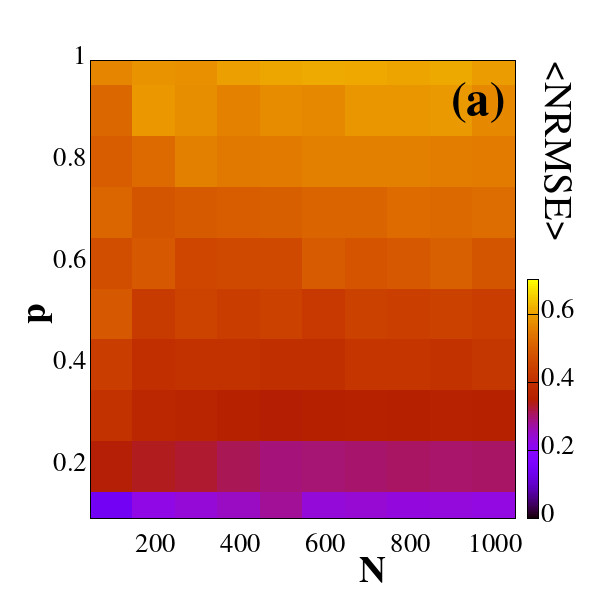}
\includegraphics[clip,width=0.45\textwidth,angle=0]{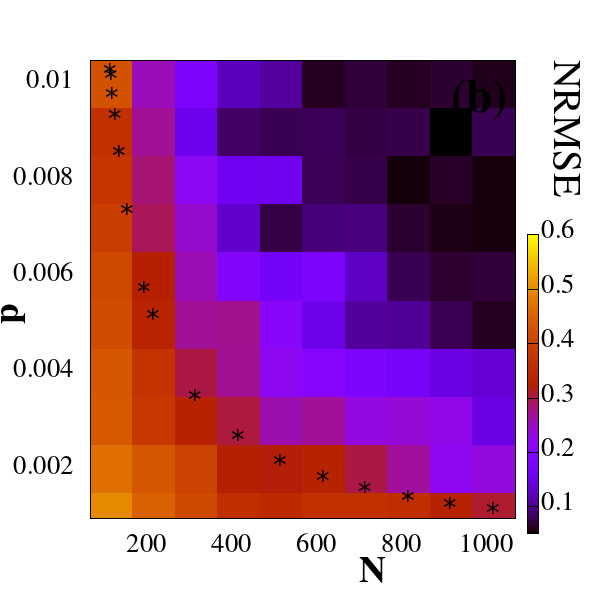}
\includegraphics[clip,width=0.45\textwidth,angle=0]{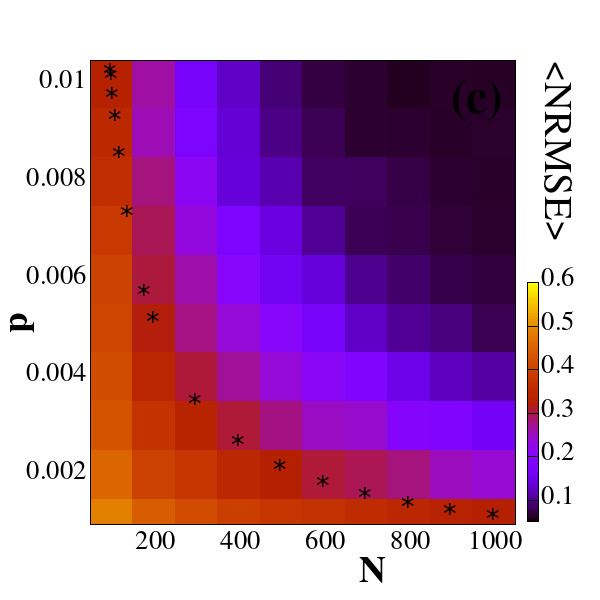}
\includegraphics[clip,width=0.45\textwidth,angle=0]{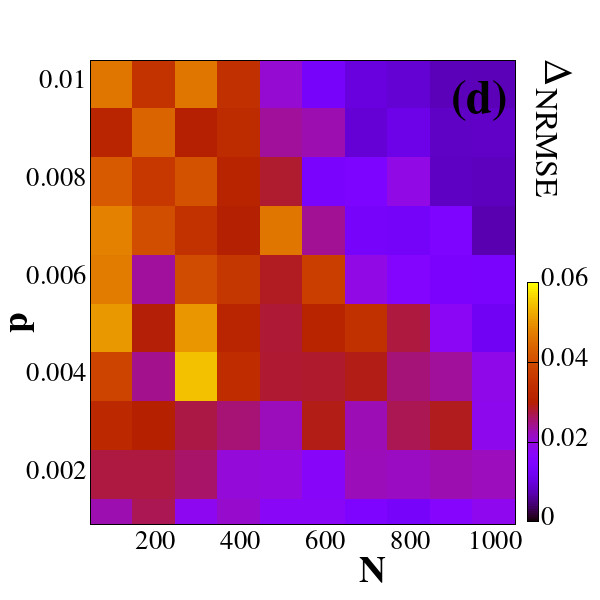}
\caption{\label{figC2} 
Lorenz chaotic dynamics with a smaller training set containing $K_{\rm training}=10,000$ data points.
All other parameters are the same as in Fig.~\ref{fig02}.
 Color-coded map of the reservoir performance (NRMSE) is presented
as a function of the reservoir size $N$ and
the Erd\"os-R\'enyi reservoir connectivity $p$. (a) Average reservoir
performance over 20 ER network realizations, for the case of large values of $p \in \left[ 0.1 , 1 \right] $,
 (b) the case of smaller $p$-values (sparse reservoir), $p \in \left[ 0.001 , 0.01 \right] $ with one ER reservoir realization,
(c) the average reservoir performance ($\left< {\rm NRMSE} \right>$) over 20 realizations of the
ER network, with $p \in \left[ 0.001 , 0.01 \right] $, and (d) the performance deviation ($\Delta_{\rm NRMSE}$) over the 20 realizations.
In panels (b) and (c) the stars denote the critical point values of mean-field bond percolation. 
}
\end{figure}

 To explore this effect in other parameter regions, in Fig.~\ref{figC2}
we plot the NRMSE performance using the smaller training set as a function of $N$ and $p$.
We note that the results are comparable 
with the ones of the larger set (30,000 data points) shown in Fig.~\ref{fig02}.
The worst NRMSE values reach the values of 0.5-0.6 for both sizes of the training sets.
However, the best predictions in the case of 10,000 points are shifted to the right with
respect to the use of 30,000 data points.
This means that for equally good predictions we need to use larger reservoirs
if the training data size is smaller.
Qualitatively, though,the best predictions are obtained beyond
 the starred line denoting the critical 
mean-field bond percolation curve, indicating that the reservoir connections need to cross
through the entire network in order to provide good predictions, regardless of the size
of the training set.

\clearpage

{\bf{\it{- Erd\"os-R\'enyi reservoir predicting FitzHugh-Nagumo dynamics}}}

\par
\begin{figure}[h]
\includegraphics[clip,width=0.45\textwidth,angle=0]{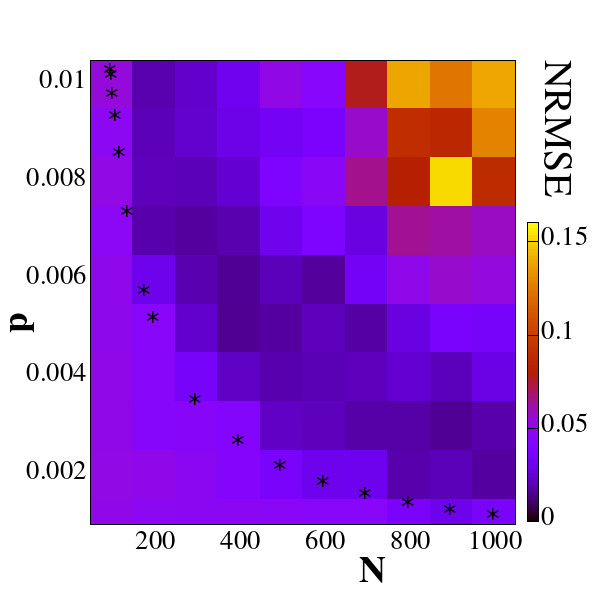}
\includegraphics[clip,width=0.45\textwidth,angle=0]{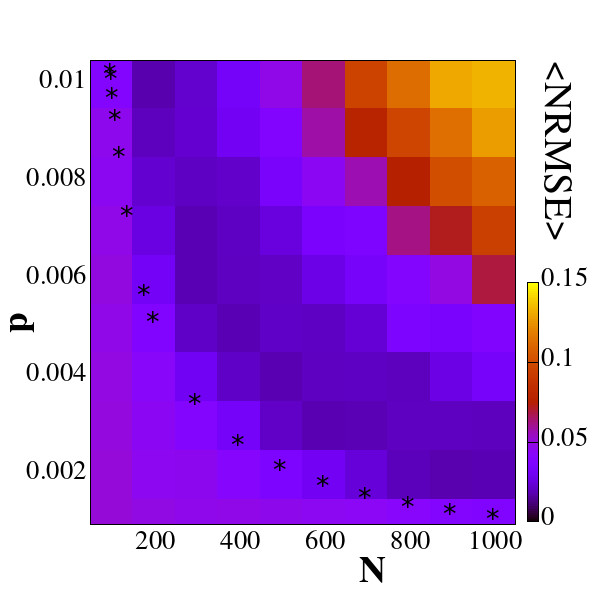}
\caption{\label{figC3} 
FitzHugh-Nagumo limit cycle dynamics; predictions with smaller training set.
 Color-coded maps of the reservoir performance (NRMSE) as a function of 
the reservoir size $N$ and the Erd\"os-R\'enyi reservoir connectivity $p$.  
(a) The NRMSE values using one ER network realization, and (b) Average $\left< {\rm NRMSE} \right>$ 
over 20 ER
network realizations.
The star points denote the critical point values of the mean-field bond percolation.
Parameter $K_{\rm training}=10,000$ and all other parameters are as in Fig.~\ref{fig04}.
 The parameters
of the FHN model data are provided in Sec.~\ref{sec:FHN}.
}
\end{figure}

  For the case of the FHN dynamics we present here the reservoir performance results
using a smaller training data set containing 10,000 points.
All other parameters are as in Fig.~\ref{fig04}. The performance results are plotted in Fig.~\ref{figC3}.

\par We note that the reservoir performances using the smaller training set (10,000 data points) 
are comparable 
with the ones of the larger set (30,000 data points) shown in Fig.~\ref{fig04}. One
characteristic difference is that the worst  NRMSE values for the smaller data set reach 0.15, while
using 30,000 data points the worst NRMSEs reach 0.12 . Such a small improvement is insufficient 
to justify the increased computational cost associated with a much larger training set.
Comparative qualitative conclusions are drawn from the smaller training set: best predictions
are obtained near the critical bond mean-field percolation curve and to the 
right of this line.

\clearpage
{\bf{\it{ - Watts \& Strogatz reservoir predicting Lorenz dynamics}}}
\par
\begin{figure}[h]
\includegraphics[clip,width=0.45\textwidth,angle=0]{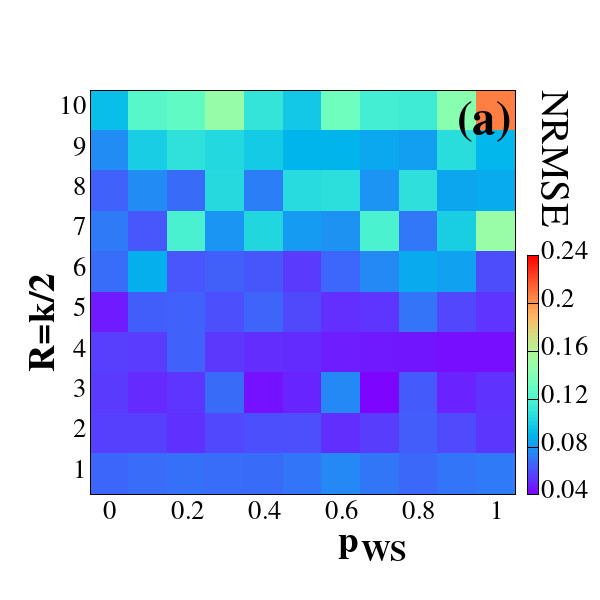}
\includegraphics[clip,width=0.45\textwidth,angle=0]{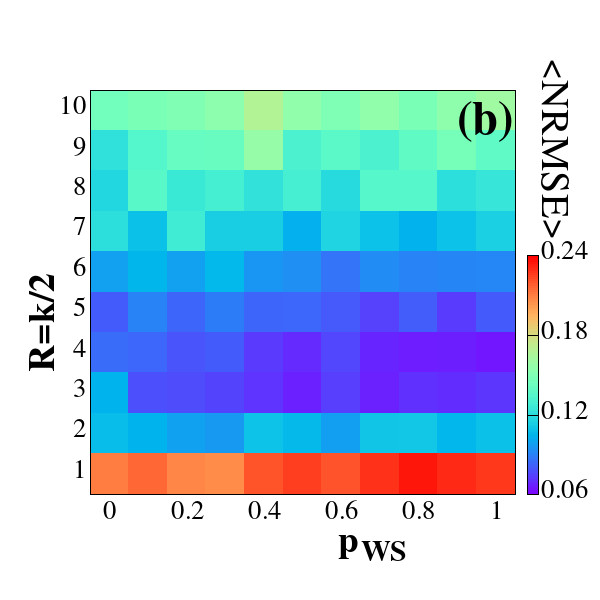}
\caption{\label{figC4} 
Lorenz chaotic dynamics predicted by a WS reservoir with smaller training set.
 Color-coded maps of the reservoir performance (NRMSE) as a function of 
the reservoir (ring) coupling range $R$ and the rewiring probability $p_{WS}$. 
(a) The NRMSE values using one WS network realization, and (b) Average $\left< {\rm NRMSE} \right>$ 
over 20 ER
network realizations.
Parameter $K_{\rm training}=10,000$ and all other parameters are as in Fig.~\ref{fig05}.
 The parameters
of the Lorenz model data are provided in Sec.~\ref{sec:Lorenz}.
}
\end{figure}

  For the case of the WS reservoir and its predictive power on the Lorenz model, we present 
in Fig.~\ref{figC4} the NRMSE results
using the smaller training data set containing 10,000 points.
All other parameters are as in Fig.~\ref{fig05}. 

\par We  note that the reservoir performances using the smaller training set (also composed by 10,000 data points) 
slightly deteriorate as compared to
the ones of the larger set (30,000 data points) shown in Fig.~\ref{fig05}.
While in the former larger set the NRMSEs were ranging between $0.04 - 0.16$, for the smaller
set here the range is relatively higher $0.06 - 0.24$.  
These conclusions are similar to the previous cases of the ER reservoirs: larger training sets
lead to better reservoir performances.
With respect to the role of the coupling range $R$ in the WS reservoir,
 the same conclusions are drawn from 
both the larger and the smaller training set: best predictions
are obtained for the smaller coupling range, $R$, values. 

\end{document}